\documentclass[reprint,aps,prx,groupedaddress,amsmath,amssymb,superscriptaddress]{revtex4-2}
\usepackage{graphicx}% Include figure files
\usepackage{dcolumn}% Align table columns on decimal point
\usepackage{bm}% bold math
\usepackage{xcolor}
\usepackage[normalem]{ulem}
\usepackage[USenglish]{babel}
\usepackage{float}
\usepackage{times}
\usepackage{physics}
\usepackage[colorlinks]{hyperref}

\newcommand*\dbcurl{\vcenter{\hbox{\includegraphics[scale=0.08]{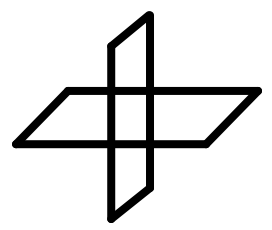}}}}

\newcommand*\gradAsq{\vcenter{\hbox{\includegraphics[scale=0.08]{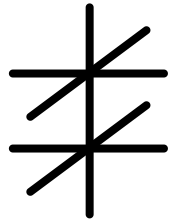}}}}

\newcommand*\pressureop{\vcenter{\hbox{\includegraphics[scale=0.08]{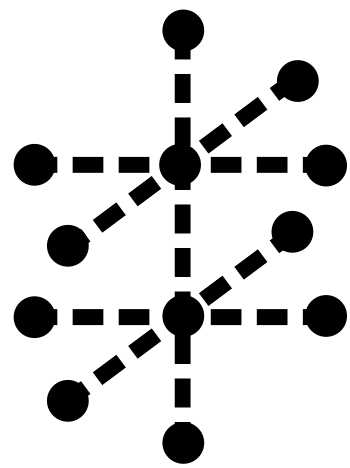}}}}

\global\long\def\bkt#1{\left(#1\right)}
\global\long\def\sbkt#1{\left[#1\right]}

\global\long\def\re{\mathrm{Re}}
\global\long\def\im{\mathrm{Im}}

\global\long\def\mb#1{{\mathbf #1}}

\makeatletter
\let\ORIbbl@fixname\bbl@fixname
\def\bbl@fixname#1{%
  \@ifundefined{languagealias@\expandafter\string#1}
    {\ORIbbl@fixname#1}
    {\edef\languagename{\@nameuse{languagealias@#1}}}%
}
\newcommand{\definelanguagealias}[2]{%
  \@namedef{languagealias@#1}{#2}%
}
\makeatother

\definelanguagealias{en}{english}
\definelanguagealias{EN}{english}

\begin{document}

% \preprint{AIP/123-QED}

%\title[Sample Title]{Sample Title:\\with Forced Linebreak\footnote{Error!}}% Force line breaks with \\
%\thanks{Footnote to title of article.}

\title[]{DEC-QED: A flux-based 3D electrodynamic modeling approach to superconducting circuits and materials}
%\thanks{Footnote to title of article.}

\makeatletter
%\long\def\@makefntext#1{%
%  \parindent 1em\noindent \hb@xt@ 1.8em{\hss \textsuperscript(\kern-0.1ex\@makefnmark\kern-0.1ex\textsuperscript)}#1}
% \let\@fnsymbol\@fnsymbol@latex
% \@booleanfalse\altaffilletter@sw
% \makeatother

\date{\today}% It is always \today, today,
             %  but any date may be explicitly specified

\author{Dung N. Pham}
\affiliation{Department of Electrical and Computer Engineering, Princeton University, Princeton, NJ 08544, USA}
\author{Wentao Fan}
\affiliation{Department of Physics, Princeton University, Princeton, NJ 08544, USA}
% \author{Kanupriya Sinha}
\author{Michael G. Scheer}
\affiliation{Department of Physics, Princeton University, Princeton, NJ 08544, USA}
\author{Hakan E. T\"ureci}
% \email{tureci@princeton.edu}
\affiliation{Department of Electrical and Computer Engineering, Princeton University, Princeton, NJ 08544, USA}
\begin{abstract}
Modeling the behavior of superconducting electronic circuits containing Josephson junctions is crucial for the design of superconducting  information processors and devices. In this paper, we introduce DEC-QED, a computational approach for modeling the electrodynamics of superconducting electronic circuits containing Josephson junctions in arbitrary three-dimensional electromagnetic environments. DEC-QED captures the non-linear response and induced currents in BCS superconductors and accurately captures phenomena such as the Meissner effect,  flux quantization and Josephson effects. Using a spatial coarse-graining formulation based on Discrete Exterior Calculus (DEC), DEC-QED can accurately simulate transient and long-time dynamics in superconductors. The expression of the entire electrodynamic problem in terms of the gauge-invariant flux field and charges makes the resulting classical field theory suitable for second quantization.
\end{abstract}

\maketitle

%\pacs{TBD}% PACS, the Physics and Astronomy
                             % Classification Scheme.
%\keywords{TBD}%Use showkeys class option if keyword
                              %display desired

\section{Introduction}
Accurate modeling of superconducting microwave circuits incorporating non-linear Josephson junction (JJ) based elements is essential for the design, control, and deployment of quantum information processing systems involving qubits and their readout systems. As the number of qubits increases and their electromagnetic environments become more complex~\cite{packaging1_2021}, efficient computational approaches are required to produce reduced quantum models that capture the relevant degrees of freedom. This active research area~\cite{BBQ_2012, EPR, JJ_embeddedTL_2012, BBQ_exact_impedance_2014, quantize_shunted_sc_2016, 1+1D_Moein_2016, cuttoff_free_cqed_2017, Anlage_GL, DiVincenzo_timedependent_magneticfield, parra_Rodriguez_2019, scq_beyond_dispersive_2019, Michael_decaymodeling, circuitQ}, which lies at the intersection of computational electromagnetism and quantum electrodynamics of superconducting devices, is of great importance not only for quantum information processing, but also for a range of other applications, including low-noise amplification, large-format detector arrays for astrophysics, and fast digital circuits.

In this paper we present DEC-QED, a computational approach for modeling superconducting non-linear elements in complex three-dimensional electromagnetic environments. DEC-QED accurately solves Maxwell's equations coupled to the non-linear Schr\"odinger equation describing the dynamics of the order parameter of the electronic condensate field of a superconductor. It utilizes the gauge-invariant flux field to describe the interactions between electromagnetic fields and charge degrees of freedom in superconducting materials. Furthermore, we demonstrate that the use of discrete differential forms and their exterior calculus, known as Discrete Exterior Calculus (DEC)~\cite{DEC_HiraniThesis, discretediffgeom_book, Chen_2017}, enables the efficient and accurate solution of the numerical problem through spatial coarse-graining. We illustrate the capabilities of DEC-QED through examples including the simulation of the Meissner effect, flux quantization, and Josephson oscillations.

The flux-field description has been essential in the development of circuit quantum electrodynamics (cQED)~\cite{blaisrmp}, a formal quantum electrodynamic theory of light-matter interactions involving macroscopic quantum degrees of freedom acting as atoms. This theory has been successfully implemented through computational approaches that synthesize accurate low-energy quantum Hamiltonians based on the flux-field description~\cite{BBQ_2012}. In this paper, we demonstrate that the flux-field description emerges naturally from a coarse-grained formulation of the electromagnetic and charge degrees of freedom. This shift in perspective is significant because it demands accuracy not at the point-wise level, but for averages of the fields over spatial intervals. DEC provides a natural framework for coarse-grained fields, enabling the accurate construction of differential operators and non-linear terms in the wave equation. We also show that under the appropriate geometric constraints, the 3+1D theory reduces to the standard 1+1D flux-field description of cQED for a transmission line~\cite{blaisrmp, 1+1D_Moein_2016} and is able to capture the known non-linear dynamics of the gauge-invariant phase of a JJ. The theory provides an ab-initio parametrization of the reduced equations and corrections to the known non-linear JJ dynamics.

As quantum processors become more complex, there are a number of computational and fundamental challenges that need to be addressed. From a practical perspective, there are challenges related to transient currents within superconductors and unintended electromagnetic interactions between circuit elements, known as cross-talk~\cite{Crosstalk1_2011, Crosstalk2_2016, Crosstalk3_2020, surfaceloss1_2015, surfaceloss2_2016}. These challenges can be addressed at the semiclassical level by describing the superconductor using its order parameter equations and the electromagnetic degrees of freedom using the classical Maxwell's equations. The focus of this paper is on computational problems of this nature. There are also fundamental challenges related to eliminating certain degrees of freedom in order to arrive at a reduced quantum model~\cite{1+1D_Moein_2016, cuttoff_free_cqed_2017, kanupaper}. The efficient computational description of various damping and decoherence effects, such as spontaneous emission and the Purcell effect~\cite{PurcellDecay_Houck_2008}, as well as quasi-particle related damping and decoherence~\cite{quasiparticles_Grunhaupt2018, quasiparticles_Catelani2011, quasiparticles_Wang2014} in the presence of non-linearity and multi-mode coupling, is a broader problem that will be addressed in future work.

The aim of this work is to rigorously extend the sub-gap electrodynamics of superconducting electrical circuits to three-dimensional structures that include JJs. To achieve this, we will (1) extend the flux-field description to 3+1D in a rigorous manner, (2) accurately account for processes occurring within superconductors rather than bypassing them using boundary conditions or the London theory, (3) capture the time dynamics of both material and microwave degrees of freedom, and (4) use a formulation that is manifestly gauge-invariant, achieved through the use of hybridized field degrees of freedom. %(5) A spectral expansion that utilizes exact modes of the open system, (6) Quantization based on (5), and (7) A multi-scale quantum perturbation theory to address the non-linearity. + (8) Introduction of a computationally accurate and efficient dual-lattice description that utilizes spatio-temporally coarse-grained fields.  (10) Do this in a way that we can recover the 1+1D circuit quantum electrodynamics under an appropriate limiting procedure. 

The paper is structured as follows. In Section \ref{sec:formulation}, we introduce the fundamental equations and field variables that will be utilized to study the semi-classical dynamics of the condensate field in superconducting materials embedded in a three-dimensional electromagnetic environment. The discretized equations for the spatially coarse-grained fields, derived using the DEC methodology, are presented in Section \ref{sec:discrete_EM_formulation}. The numerical results of our analysis, based on these equations, are presented in Section \ref{sec:numerics}. Specifically, in Section \ref{sec:linearmodes}, we discuss the eigenmodes of the resulting vector Helmholtz equation obtained through the application of DEC. The time-domain simulation of an oscillating dipole inside a superconducting cavity with finite-width walls is discussed in Section \ref{sec:oscillating_dipole}. The simulation of the dynamical Meissner effect is presented in Section \ref{sec:meissner_effect}, while the investigation of flux quantization in superconducting loops is presented in Section \ref{sec:flux_quantization}. Finally, the ab-initio modeling of Josephson junction dynamics is discussed in Section \ref{sec:JJdynamics}.

\section{Semi-classical gauge-invariant formulation of electrodynamics in superconducting materials}\label{sec:formulation}

% \ks{general suggestion - some details of the derivation from this section  can be skipped/moved to Appendix. }
\begin{figure}[h!]
    \centering
    \includegraphics[scale=0.37]{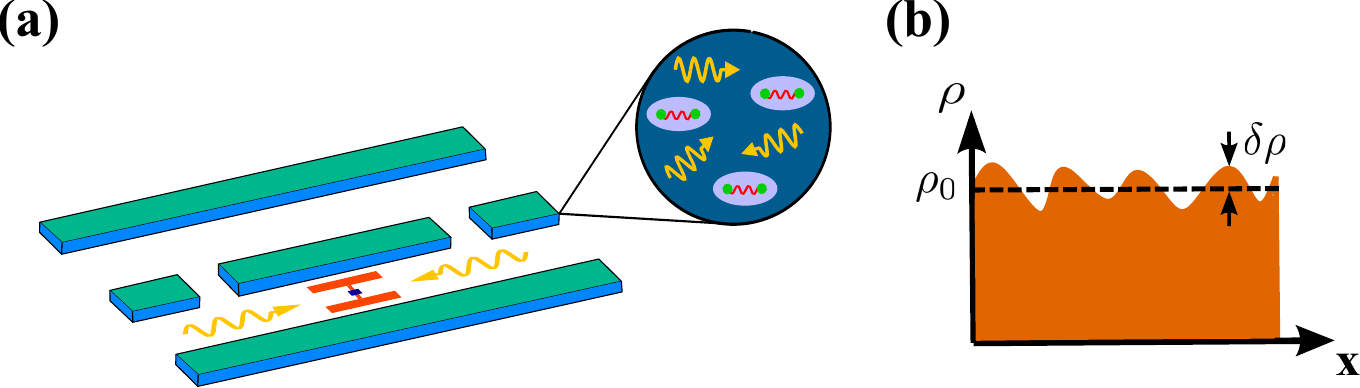}
    \caption{(a) A schematic of a superconducting material interacting with the electromagnetic field~\cite{blaisrmp}. The example used here is a superconducting JJ-based qubit embedded in a coplanar waveguide resonator. (b) An example of the spatial dependence of the density of condensate electrons $\rho$ in the jellium model, in which the charge fluctuation $\delta\rho$ on top of the mean value $\rho_0$ is small, i.e. $\delta\rho\ll\rho_0$.}
    \label{fig:model_schem}
\end{figure}

We consider a superconducting material consisting of potentially multiple disconnected domains interacting with the electromagnetic (EM) field in a three dimensional volume (see Fig.\,(\ref{fig:model_schem})). A Jellium-like model is used to describe the superconducting material, in which the solid provides an immobile ionic continuum that exactly balances the total charge of the dynamical superconducting condensate field. The dynamics of the order parameter $\Psi\bkt{{\bf r},t}$ describing the condensate electrons is then given by \cite{Feynman_lectures}
\begin{equation}\label{eq:generalSE}
    i\hbar\frac{\partial \Psi({\bf r},t)}{\partial t} 
    = \bigg[\frac{1}{2m}\Big(\!\!-i\hbar{\bf \nabla}-q{\bf A} \Big)^2 + qV({\bf r},t) + U({\bf r}) \bigg]\Psi\bkt{{\bf r},t}
\end{equation}
where $m=2m_e$ and $q=-2e$ correspond to the mass and charge of a superconducting Cooper pair respectively, ${\bf A}$ is the magnetic vector potential, $V$ is the scalar electric potential of the condensate charge field, and $U$ is the static potential defined by the superconducting material. The spatial-dependence of the potential $U$ will be used to define different materials (and vacuum) that confine the condensate field.  If there is only one material in consideration, then $U$ is constant. In the Madelung representation~\cite{madelungpaper} for the condensate wavefunction, we write $\Psi({\bf r},t) = \sqrt{\rho({\bf r},t)}e^{i\theta({\bf r},t)}$, where $\rho$ and $\theta$ are the density and phase of the condensate, respectively. The supercurrent is then given by 
\begin{equation}\label{eq:supercurrent_def}
{\bf J}_s = \frac{q}{m}\re\bigg\{\Psi^*\Big(\frac{\hbar}{i} \nabla-q{\bf A}\Big)\Psi\bigg\} = \frac{q\rho}{m}\Big(\hbar\nabla\theta - q{\bf A} \Big).
\end{equation}
Rewriting Eq.\,(\ref{eq:generalSE}) using the Madelung representation we obtain from the imaginary part of Eq.\,(\ref{eq:generalSE}) the current continuity equation
\begin{align}\label{eq:rho_EoM}
\frac{\partial\rho}{\partial t} = -\frac{1}{q}\nabla \cdot {\bf J}_s,
\end{align}
while the real part of Eq.\,(\ref{eq:generalSE}) gives the equation of motion for the phase $\theta$
\begin{align}\label{eq:theta_EOM2}
\frac{\partial \theta}{\partial t} = -\frac{m}{2\hbar q^2\rho^2}J^2_s + \frac{\hbar}{2m}\frac{\nabla^2(\sqrt{\rho})}{\sqrt{\rho}} - \frac{1}{\hbar}(qV + U).
\end{align}
%The third term in RHS of Eq.\,(\ref{eq:theta_EoM}) negligible compared to other terms, and the equation of motion for $\theta$ in a uniform material is given by
%\begin{equation}\label{eq:theta_EOM2}
%    \frac{\partial\theta}{\partial t} \approx -\frac{q^2}{2m\hbar}\bigg|\frac{\hbar}{q}\nabla\theta - {\bf A} \bigg|^2 - \frac{qV}{\hbar}.
%\end{equation}
The dynamics of the vector and scalar potentials ${\bf A}$ and $V$ are governed by the Maxwell's equations as
\begin{align}
    {\bf\nabla}\times{\bf\nabla}\times{\bf A} + \mu_0\epsilon_0\ddot{{\bf A}} &= \mu_0({\bf J}_s + {\bf J}_{\text{src}}) - \mu_0\epsilon_0\nabla\dot{V}, \label{eq:Ampere} \\
    -\nabla^2V &= \frac{q}{\epsilon}(\rho +\rho_{\text{src}}), \label{eq:Gauss}
\end{align}
where $\rho_{\text{src}}$ and ${\bf J}_{\text{src}}$ are external charge and current sources. Next, we introduce the gauge-independent hybridized field
\begin{equation}
    {\bf A}' = {\bf A} - \frac{\hbar}{q}\nabla\theta.
    \label{eq:hybA}
\end{equation}
Using Eq.\,(\ref{eq:supercurrent_def}) to substitute ${\bf J}_s$, Eq.\,(\ref{eq:Ampere}) can be rewritten in terms of the new field ${\bf A}'$ and $\rho$ as  
\begin{align}\label{eq:Aprime_waveeq_rho}
 {\bf\nabla}\!\times\!{\bf\nabla}\!\times\!{\bf A'} + \mu_0 &\epsilon_0\frac{\partial^2{\bf A'}}{\partial t^2} + \frac{\mu_0 q^2}{m}\rho{\bf A'}  -\frac{\mu_0\epsilon_0 q}{2m}\frac{\partial}{\partial t}\nabla\big|{\bf A'}\big|^2 \nonumber \\
  & + \frac{\mu_0\epsilon_0\hbar^2}{2mq}\frac{\partial}{\partial t}\nabla\bigg[\frac{\nabla^2(\sqrt{\rho})}{\sqrt{\rho}}\bigg] =  \mu_0{\bf J}_{src} .
\end{align}
%where $\lambda_L = \sqrt{\frac{m}{\mu\rho_0q^2}}$ is the London penetration depth in zero field. 
Eq.\,(\ref{eq:rho_EoM}) can also be written explicitly as
\begin{equation}\label{eq:chargeconserve2_rho}
    \frac{\partial\rho}{\partial t}  = {\bf \nabla} \cdot \Bigg[\frac{q}{m}\rho{\bf A'} - \frac{{\bf J}_{src}}{q}\Bigg] - \frac{\partial\rho_{src}}{\partial t}.
\end{equation}
Eqs.\,{(\ref{eq:Aprime_waveeq_rho}), (\ref{eq:chargeconserve2_rho})} are the two central equations of our formulation, with which one can study the time-evolution of systems of piece-wise superconducting objects embedded in the three-dimensional vacuum and interacting with an electromagnetic environment. The hybridized field ${\bf A}'$ spans the entire space and lives both inside the materials and in vacuum. 

For macroscopic superconducting devices, it is reasonable to assume that the electron condensate is nearly ``rigid" in the bulk, meaning that the time-dependent electromagnetic (EM) field only causes a small fluctuation in the distribution of Cooper pairs. As a result, in the bulk of the material, the charge density can be expressed as $\rho=\rho_0+\delta\rho$, where $\rho_0$ is the uniform, static charge distribution that cancels out the charge from the ionic lattice in the unperturbed case and $\delta\rho$ is the fluctuation due to the presence of EM fields. Eqs.~(\ref{eq:Aprime_waveeq_rho}) and (\ref{eq:chargeconserve2_rho}) can be solved without this approximation; the equations we will be solving below are non-linear as well. This approximation allows however for the introduction of important physical scales in a natural way and enables a systematic expansion in $\delta\rho/\rho_0$ if greater accuracy is required. As demonstrated in the numerical results of section \ref{sec:numerics}, when combined with the coarse-grained formulation provided by DEC, this approximation can be reliably extended to the boundaries of macroscopic objects.

%In the case of macroscopic superconducting devices, it is safe to assume that the electron condensate is almost ``rigid" in the bulk, in the sense that the time-dependent EM field only creates a small fluctuation in the distribution of Cooper pairs.  Therefore in the bulk of the material $\rho= \rho_0 + \delta\rho$, with $\delta\rho\ll\rho_0$, where $\rho_0$ is the uniform, static charge distribution that in the unperturbed case cancels out the charge coming from the ionic lattice with same density, and $\delta\rho$ is the fluctuation due the presence of EM fields. Eqs.~(\ref{eq:Aprime_waveeq_rho}) and (\ref{eq:chargeconserve2_rho}) can be solved without this approximation; the equations we will be solving below are non-linear as well. This approximation however provides a setting to introduce important physical scales in a natural way and allows for a systematic expansion in $\delta \rho/\rho_0$ if more accuracy is demanded. As will be shown in the numerical results in Sec.\,\ref{sec:numerics}, when combined with the coarse-grained formulation provided by DEC, this approximation can be reliably extended to the boundaries of macroscopic objects.

Introducing the penetration depth of a superconductor $\lambda_L = \sqrt{\frac{m}{\mu\rho_0q^2}}$, Eq.\,(\ref{eq:Aprime_waveeq_rho}) can be decomposed into linearized and nonlinear parts. Before we do that, notice that the last term in the LHS of Eq.\,(\ref{eq:Aprime_waveeq_rho}), the quantum pressure term, can be shown to be very small in the bulk of the superconductor. In fact, expanding $\sqrt{\rho} \approx \sqrt{\rho_0}(1+\delta\rho/2\rho_0)$ and using Eq.(\ref{eq:chargeconserve2_rho}), one can show that this term becomes  $\frac{\mu\epsilon\hbar^2}{4m^2}\nabla\big(\nabla^2({\bf \nabla}.{\bf A}')\big)$. The smallness parameter for this term is $\frac{\hbar^2}{4m^2c^2\lambda_L^4}\approx 10^{-12}$ inside the superconductor for $\lambda_L=100$\,nm, justifying our neglecting this term. 
Eqs.\,(\ref{eq:Aprime_waveeq_rho}) and (\ref{eq:chargeconserve2_rho}) can then be written as
\begin{align}\label{eq:Aprime_waveeq}
 {\bf \nabla}\!\times\!{\bf \nabla}\!\times\!{\bf A'} + \mu_0\epsilon_0&\frac{\partial^2{\bf A'}}{\partial t^2} + \frac{1}{\lambda_L^2}{\bf A}' + \frac{\mu_0 q^2}{m}\delta\rho{\bf A'} \nonumber \\
  & -\frac{\mu_0\epsilon_0 q}{2m}\frac{\partial}{\partial t}\nabla\big|{\bf A'}\big|^2 =  \mu_0{\bf J}_{src} . 
\end{align}
and
\begin{equation}\label{eq:chargeconserve2}
    \frac{\partial\delta\rho}{\partial t}  = {\bf \nabla}\cdot\Bigg[\bigg(\frac{1}{\mu_0 q\lambda_L^2} + \frac{q\delta\rho}{m} \bigg){\bf A'} - \frac{{\bf J}_{src}}{q}\Bigg] - \frac{\partial\rho_{src}}{\partial t},
\end{equation}
Equations (\ref{eq:Aprime_waveeq}) and (\ref{eq:chargeconserve2}) accurately describe the evolution of ${\bf A}'$ and $\delta\rho$ {\it in the bulk} of the superconducting material. The first three terms on the left-hand side of equation (\ref{eq:Aprime_waveeq}) constitute the linear response of a superconductor as described by London theory~\cite{londontheory}, while the remaining terms are non-linear corrections. Given the source terms, these equations form a complete set of equations. However, near the boundaries, it is necessary to include non-linear corrections from the quantum pressure term. Section \ref{sec:JJdynamics} provides a detailed numerical analysis and analytical derivations where such boundary terms become important, solving the full equations Eqs.\,(\ref{eq:Aprime_waveeq_rho}) and (\ref{eq:chargeconserve2_rho}), or their discretized version, Eqs.\,(\ref{Eq:DiscreteAmpereLaw_full}) and (\ref{Eq:DiscreteChargeEoM}).

%These equations are very accurate in the bulk of the superconducting material. In Eq.\,(\ref{eq:Aprime_waveeq}) the first three terms constitute the linear response of a superconductor given by the London theory~\cite{londontheory} while the remaining terms on the LHS are the non-linear corrections. Given the source terms, Eqs.\,(\ref{eq:Aprime_waveeq}) and (\ref{eq:chargeconserve2}) form a complete set of equations that describes the evolution of ${\bf A}'$ and $\delta\rho$. Near the boundaries however, the non-linear corrections from the quantum pressure term have to be included. Sec.~\ref{sec:JJdynamics} provides a detailed numerical analysis and analytical derivations where such boundary terms become important. 

Note that the ${\bf A}'$ field here is a light-matter field containing both  electromagnetic (${\bf A}$) and the condensate ($\nabla\theta$) degrees of freedom. Using such a hybrid field in the formulation helps us efficiently capture the dynamics of both the EM field and the material and their interactions. Indeed, the non-linearities in Eq.\,(\ref{eq:Aprime_waveeq}) arise from such interactions: the term $\delta\rho{\bf A}'$ comes from the fluctuation in the supercurrent inside the material due to a time-dependent EM field, while the term $\nabla|{\bf A}'|^2$ derives from the nonlinear dependence of the superconducting phase $\theta$ on the EM field (Eq.\,(\ref{eq:theta_EOM2})).

%As an alternative to Eqs.\,(\ref{eq:Aprime_waveeq}) and (\ref{eq:chargeconserve2}), we can use Gauss's law to rewrite the hybridized light-matter dynamics of ${\bf A}'$ and $\nabla V$ as
%\begin{align}\label{eq:Aprime_waveeq2}
%   {\bf \nabla}\!\times\!{\bf \nabla}\!\times\!{\bf A'} + \mu\epsilon\frac{\partial^2{\bf A'}}{\partial t^2} + &\frac{1}{\lambda_L^2}{\bf A'}\!-\!\frac{\mu\epsilon q}{m}\nabla^2V{\bf A'}  - \frac{\mu q^2}{m}\rho_{src}{\bf A'} \nonumber \\
%  & - \mu{\bf J}_{src} -\frac{\mu\epsilon q}{2m}\frac{\partial}{\partial t}\nabla\big|{\bf A'}\big|^2 = 0,
%\end{align}
%and
%\begin{equation}\label{eq:gausslaw2}
%    \nabla^2\dot{V} = -{\bf \nabla}.\Bigg\{\bigg[\frac{1}{\epsilon\mu \lambda_L^2} - \frac{q}{m}\nabla^2V \bigg]{\bf A'} \Bigg\} - \frac{q}{\epsilon}\dot{\rho}_{src}.
%\end{equation}
%The hybridized light-mater dynamics in a superconducting material can then be simulated by numerically solving either Eqs.\,(\ref{eq:Aprime_waveeq}) and (\ref{eq:chargeconserve2}) or Eqs.\,(\ref{eq:Aprime_waveeq2}) and (\ref{eq:gausslaw2}). For completeness, in Appendix \ref{append:perturbationtheory} we also provide a perturbative analysis to treat the non-linearities in Eqs.\,(\ref{eq:Aprime_waveeq}) and (\ref{eq:chargeconserve2}).
The hybridized light-mater dynamics in a superconducting material can then be simulated by numerically solving either Eqs.~(\ref{eq:Aprime_waveeq_rho}) and (\ref{eq:chargeconserve2_rho}) or the pair Eqs.\,(\ref{eq:Aprime_waveeq}) and (\ref{eq:chargeconserve2}). For completeness, in Appendix \ref{append:perturbationtheory} we also provide a perturbative analysis to treat the non-linear terms in Eqs.\,(\ref{eq:Aprime_waveeq}) and (\ref{eq:chargeconserve2}).

\section{Electrodynamics with coarse-grained flux variables}\label{sec:discrete_EM_formulation}
\subsection{Discretization of Maxwell's electrodynamics}
\begin{figure*}[t]
    \centering
    \includegraphics[scale=0.23]{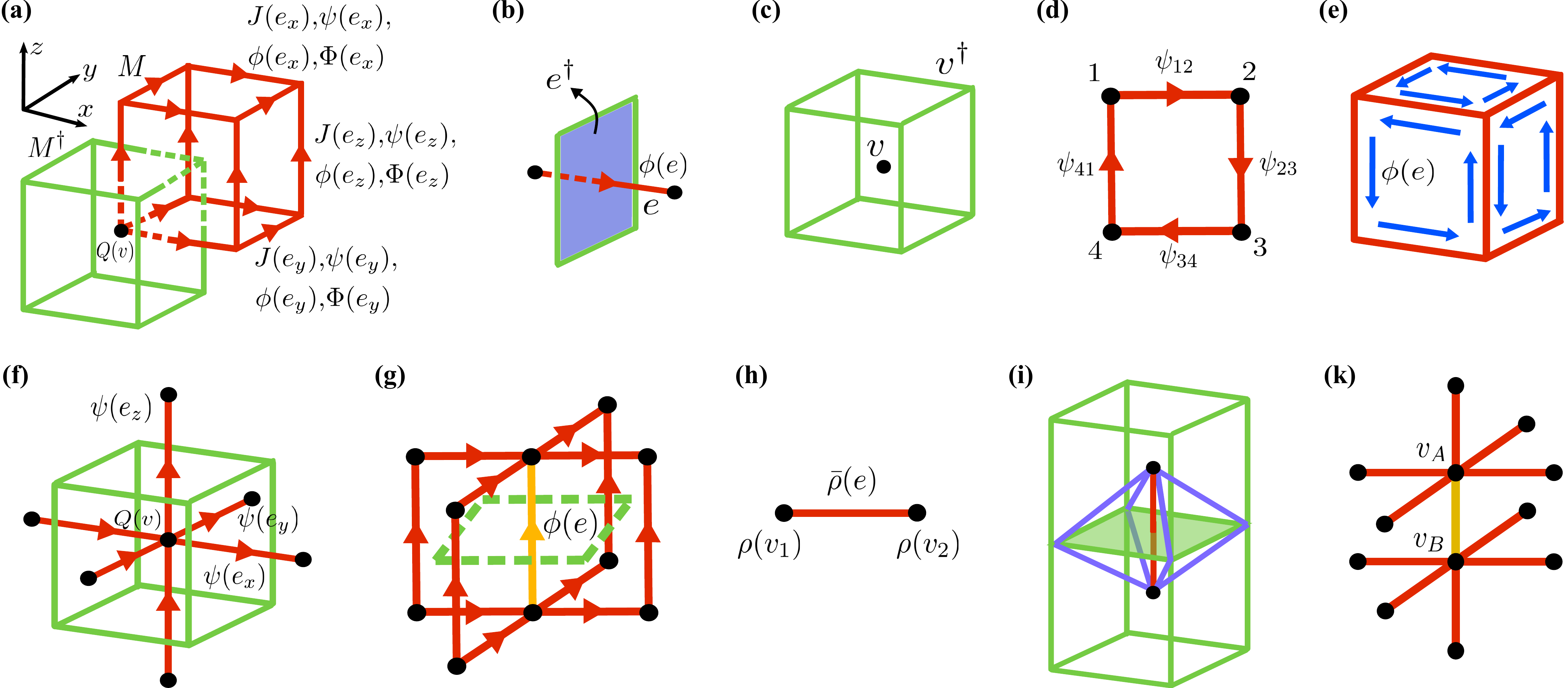}
    \caption{(a) A dual-lattice cubical mesh is shown, in which the edges of the primal lattice $M$ are in red, while the edges of the dual lattice $M^{\dagger}$ are in green. The fields $\psi,\phi,\Phi$ and the current density ${\bf J}$ are defined on the edges. (b) An edge $e$ in $M$ is orthogonal to a corresponding face $e^{\dagger}$ in $M^{\dagger}$. (c) A vertex $v$ in the primal mesh corresponds to a cell $v^\dagger$ in the dual mesh. 
    (d) Illustration of how the identity ${\bf \nabla}\times(\nabla V)=0$ is naturally satisfied on the boundary of each face. (e) Similarly, the identity ${\bf \nabla}.({\bf \nabla}\times{\bf A})=0$ is also locally satisfied at each unit volume.
    (f) The divergence operator applied to a vertex $v$ corresponds to a sum (with the correct signs) of flux lines that pass through it. (g) The double curl operator $\nabla\times\nabla\times$ applied to an edge $e$ (shown in yellow) corresponds to a weighted sum of that edge with the surrounding edges in $M$. (h) The $\bar{\rho}$ field, defined on every edge $e$, is determined by averaging the values of $\rho$ evaluated at the vertices that $e$ connects. (i) An example of the support volume of a primal edge is shown. The primal edge is red, the dual lattice is green, and the support volume in purple. (k) To represent the action of the gradient term $\nabla|{\bf A}'|^2$ on an edge (colored yellow), $|{\bf A}'|^2$ needs to be defined at the two vertices $v_A$ and $v_B$. To do so, the values of $\phi(e)$ at the surrounding edges are needed, resulting in two sets of concurrent edges connected via the shared edge. The geometrical representation of pressure term is somewhat similar. The main difference is that scalar fields at vertices are in use instead of edge fields.} 
    \label{fig:dualmesh_DEC}
\end{figure*}
This section outlines the method by which the discretization of Maxwell's theory of electromagnetism is achieved through the use of DEC. This discussion aligns with previous research on DEC formulations of Maxwell's equations~\cite{Chen_2017, Chen2_2017}.  In this specific instance, we utilize a variant that is formulated in terms of the potentials $({{\bf A}, V})$ as a foundation for deriving the equations of a superconductor. Here the spatio-temporal behaviour of sources is assumed to be given. In the subsequent section, we will require these sources to be self-consistently determined through the order parameter equation (Eq.~\ref{eq:generalSE}).

We consider a mesh $M$ that spans over the domain of interest $D$ and decomposes it into polygonal (in 2D) or polyhedral (3D) cells. The DEC framework requires in addition the introduction of a dual mesh $M^{\dagger}$ whose vertices are the circumcenters of the cells of $M$ such that connecting two vertices in $M^{\dagger}$ creates an edge if and only if the corresponding cells in M share a face. Therefore, by construction there is a one-to-one mapping between vertices ($v$), edges ($e$), faces ($f$), and cells ($c$) in $M$ and cells, faces, edges, vertices in $M^{\dagger}$, respectively. In this paper we will use $\dagger$ to denote the duality of the two lattices. For example, $v^{\dagger}$ is the cell in $M^{\dagger}$ whose circumcenter is $v$, or $e^{\dagger}$ is the face in $M^{\dagger}$ that is orthogonal to $e$, etc. (see Fig.~\ref{fig:dualmesh_DEC}).
Starting with the canonical fields $\{{\bf A}, \nabla V\}$ that are used in standard electrodynamics, we define the following coarse-grained flux fields
\begin{align}
    \label{eq:phi_define}\phi(e) &= \int_e{\bf d\ell} \cdot {\bf A},  \\
    \label{eq:psi_define}\psi(e) &= \int_e{\bf d\ell} \cdot \nabla V
\end{align}
defined on the edges $e$ of the lattices. One can think of the fields $\phi(e)$ and $\psi(e)$ as averaged projections of the vector fields ${\bf A}$ and $\nabla V$ onto an edge $e$, respectively, but note that these quantities are scaled by the length of the edge. Figure \ref{fig:dualmesh_DEC}(a) illustrates the construction of a dual mesh and the locations of the flux fields. Discrete Exterior Calculus (DEC) is a discretized version of exterior calculus in which the derivative operation on each lattice is the discretized exterior derivative and the mapping between the two lattices is the discretized hodge star~\cite{discretediffgeom_book}. DEC was originally developed for simplicial meshes~\cite{DEC_HiraniThesis} and has been successful in modeling various systems of non-linear partial differential equations~\cite{DEC_NavierStokes, Chen_2017}. In this paper, we use cubic dual meshes, which provide a more visually intuitive interpretation of DEC's geometric constructions while still yielding the same discretized equations. We will focus on the aspects of DEC that are necessary for our physical problem, which is the formulation of light-matter interaction in superconductors, and avoid discussing the full mathematical structure of the theory.

Figures \ref{fig:dualmesh_DEC}(b) and \ref{fig:dualmesh_DEC}(c) demonstrate the dualities between elemental objects in the primal and dual meshes. At first glance, the dual mesh construction may seem unnecessarily complex, as traditional finite element methods only use one mesh. However, as this section will show, the interaction between the primal and dual meshes allows for compact descriptions of differential operators. The dual mesh is also useful when dealing with boundaries in systems with multiple materials. In such cases, the primal mesh conforms to the interfaces between objects, while the material properties can be defined on the dual mesh to account for effective values attached to edges on those interfaces~\cite{Chen2_2017}. Despite the complexity of the geometric constructions involved in its dual lattice formulation, the authors believe that the benefits of using DEC outweigh this complexity.

In DEC, vectorial quantities are {\it projected} onto oriented edges of the grid (as seen in Eqs.~(\ref{eq:phi_define}) and (\ref{eq:psi_define})), while scalars are attached to vertices. Since the rate of change of a scalar is a vector field, and vice versa, this inter-connectivity of primal edges with primal vertices (and their duals) leads to a natural representation of differential operators. The discretization of Maxwell's equations can therefore be done as follows. Consider a vector field ${\bf F}$ defined on the edges of the primary lattice $M$, then
\begin{align}\label{eq:div_dec}
    \int_{v^{\dagger}}(\nabla \cdot {\bf F}) \, d{\bf r}^3 &= \sum_{e|v\in e} \frac{\Delta A(e^{\dagger})}{\Delta\ell(e)}\int_e {\bf F} \cdot {\bf d\ell}, 
\end{align}
where $\Delta\ell(e)$ is the length of the edge $e$ and $\Delta A(e^{\dagger})$ is the area of its dual face $e^{\dagger}$. Similarly, for an edge $e\in M$, the curl-curl operator in the discrete flux language is given by

\begin{align}\label{eq:doublecurl_dec}
    \int_{e^{\dagger}}\! (\nabla\!\times\!\nabla\!\times {\bf F}) \cdot d{\bf a} &= \sum_{e_0\in\partial(e^{\dagger})} \frac{\Delta\ell(e_0)}{\Delta A(e_0^{\dagger})} \sum_{e_1\in\partial(e_0^\dagger)}\int_{e_1}\!{\bf F} \cdot {\bf d\ell},
\end{align}
where $\partial(e^{\dagger})$ and $\partial(e_0^\dagger)$ are the boundaries of $e^\dagger$ and $e_0^\dagger$, respectively. The graphical illustrations of Eqs.\,(\ref{eq:div_dec}) and (\ref{eq:doublecurl_dec}) are shown in Figs.\,(\ref{fig:dualmesh_DEC}f) and (\ref{fig:dualmesh_DEC}g), respectively, while detailed derivation of these equations are discussed in Appendix \ref{append:discreteMaxwell}. 

With the discrete forms of the divergence and curl-curl operators, Maxwell's equations can be written in terms of flux fields. For ${\bf F}\!=\!\nabla V$, Eq.\,(\ref{eq:div_dec}) yields the discrete version of Gauss's law
\begin{equation}\label{eq:discrete_GaussLaw}
    \sum_{e|v\in e} \frac{\Delta A(e^{\dagger})}{\Delta\ell(e)} \psi(e) = -\frac{Q(v)}{\epsilon},
\end{equation}
while substituting ${\bf F}\!=\!{\bf A}$ in Eq.\,(\ref{eq:doublecurl_dec}), we can rewrite Ampere's law in Eq.\,(\ref{eq:Ampere}) 
\begin{align}\label{eq:discrete_Ampere}
\sum_{e_0\in\partial(e^{\dagger})}\sum_{e_1\in\partial(e_0^\dagger)} &\frac{\Delta\ell(e_0)}{\Delta A(e_0^{\dagger})}\phi(e_1) + \mu\epsilon\frac{\Delta A(e^{\dagger})}{\Delta\ell(e)}\ddot{\phi}(e) \nonumber\\
    &= \mu I(e) - \mu\epsilon\frac{\Delta A(e^{\dagger})}{\Delta\ell(e)}\dot{\psi}(e).
\end{align}
Here $I(e) = (\Delta A(e^{\dagger})/\Delta\ell(e))\int_e{\bf J}.{\bf d\ell}$.
Eqs.\,(\ref{eq:discrete_GaussLaw}) and (\ref{eq:discrete_Ampere}) describes the evolution of the fields $(\phi(e),\psi(e))$, which are coarse-grained representations of $({\bf A}, \nabla V)$, under a given source current $I(e)$ and a charge distribution $Q(v)$. In addition to (\ref{eq:discrete_GaussLaw}) and (\ref{eq:discrete_Ampere}), there are two additional equations which serve as constraints on the fields and are given by 
\begin{align}
    {\bf \nabla}\times(\nabla V) = 0, \label{eq:curl_grad}\\
    {\bf \nabla} \cdot ({\bf \nabla}\times{\bf A}) = 0, \label{eq:div_curl}
\end{align}
These constraints, however, are automatically satisfied by the very construction of DEC. Indeed, we can show this by  integrating the LHS of Eq.\,(\ref{eq:curl_grad}) over the area of an elemental face $f$ on the primal mesh
\begin{align}\label{eq:curl_grad_prove}
    \int_f({\bf \nabla}\times(\nabla V)) \cdot d{\bf a} &= \sum_{e\in\partial f}\int_e\nabla V \cdot {\bf d\ell} \nonumber\\
    &= \sum_{i,j\in\partial e|e\in\partial f}\!\!\!\!\!\!\! V_i - V_j \nonumber\\
    &= 0, 
\end{align}
where $\partial f$ is the boundary of the face $f$, and $\partial e$ is the boundary of the edge $e$. A graphical illustration of Eq.\,(\ref{eq:curl_grad_prove}) is shown in Fig.\,(\ref{fig:dualmesh_DEC}d), where we can see that the value of $V_i$ at each vertex of the face is counted twice, but with opposite signs. Similarly, we can show that the constraint in Eq.\,(\ref{eq:div_curl}) is satisfied by doing a volume integral over a unit cubical cell
\begin{align}\label{eq:div_curl_prove}
    \int_c{\bf \nabla} \cdot ({\bf \nabla}\times{\bf A}) dr^3 &= \sum_{f \in \partial c}\int_f ({\bf \nabla}\times{\bf A}) \cdot d{\bf a} \nonumber\\
    &= \sum_{f \in \partial c}\sum_{e \in \partial f} \phi(e) \nonumber\\
    & = 0, 
\end{align}
where the final sum is taken over the boundaries $\partial f$ of the faces $f$ that form the boundary of $c$, as shown in Fig.\,(\ref{fig:dualmesh_DEC}e). The automatic satisfaction of Eqs.\,(\ref{eq:curl_grad})-(\ref{eq:div_curl}), which alleviates the need to impose these constraints manually, present an advantage of DEC. One may find the dual mesh construction of DEC similar to the Yee grid formulation~\cite{Yee_1966} used in finite difference time domain (FDTD). The key distinction of DEC is that here the fundamental variables are the ``small" integrals defined over edges for which the coarse-grained dynamical equations are written. This is a deviation from FDTD and standard finite element schemes as these methods attempt to model continuous differential equations by computing their continuous variables through spatial and temporal discretization. Another advantage of our chosen approach is that due to finite spatial and temporal resolutions of experimental apparatus, measured quantities are fundamentally coarse-grained. Therefore by directly formulating a coarse-grained description of the electromagnetic problem, our approach is appropriate for describing actual experimental measurements.  This formulation is also computationally efficient, as it does not require a finely discretized physical domain to accurately conform with methods that are based on continuous formulations.

\subsection{Coarse-grained formulation for the electrodynamics of superconductors}

In this section, we discuss a DEC formulation of the solution of Eqs.~(\ref{eq:Aprime_waveeq})-(\ref{eq:chargeconserve2}) that describe the electrodynamics of superconductors. To this end, we introduce the coarse-grained hybridized field $\Phi(e)$ defined as follows:
\begin{align}
    \label{eq:phip_define} \Phi(e) &= \int_e{\bf d\ell} \cdot {\bf A}'.
\end{align}
The use of this edge-based flux field in this work is partially motivated by its connection to the gauge-invariant phase difference $\varphi$ across a JJ, which serves as the foundation for the lumped-element formulation of cQED~\cite{devoret_JJphase_fundamentals_2021}:
\begin{align}\label{eq:JJvarphi_define}
    \varphi = \theta_2-\theta_1 - \frac{q}{\hbar}\int_1^2\!\!{\bf d\ell} \cdot {\bf A},
\end{align}
with $\theta_1$ and $\theta_2$ being the phases of the condensate wavefunction at the two insulator-superconductor interfaces of the junction. This is equivalent to the coarse-grained edge field $\varphi=-\frac{q}{\hbar}\int\!{\bf A}' \cdot {\bf d\ell}$ over a single edge spanning the insulator of a JJ. It is generally believed that the interaction of the JJ with its surrounding electromagnetic environment is fully encoded in its critical current and the gauge-invariant flux $\varphi$. DEC-QED generalizes this idea of coarse-grained encoding to all of space using the definitions in Eqs.~(\ref{eq:psi_define})-(\ref{eq:phip_define}). This allows us to extend the flux-based description of cQED to the three-dimensional domain of the full system, thereby enabling the generalization of techniques in one-dimensional cQED to treat superconducting circuits in 3D. Looked from the prism of DEC-QED, one may explain the success of cQED in describing JJ-based electrical circuits as follows: From the point of view of the accurately capturing the dynamics of the {\it coarse-grained fields} across the entire 3D domain, it is not necessary to resolve the microscopic details of light-matter dynamics inside the JJ.
On the other hand, should a detailed study of microscopic dynamics within a JJ be desired, or for more complicated junctions, DEC-QED can still capture the dynamics by refining mesh density inside the junction (see Sec.~\ref{sec:JJdynamics} for a detailed numerical analysis of the physics of a JJ, and how coarse-graining accurately captures results captured by finer meshes).

Another motivation for coarse-graining is the need for an effective multi-scale simulation technique. In a typical superconducting circuit, the JJs are the smallest structures, usually a few $nm$ in thickness, and hence orders smaller than the capacitor pads (that can be in the order of $mm$). The superconducting qubits are typically coupled to cavity resonators that are much larger than the qubits in size. Performing full-wave simulations of such a system are resource-intensive, especially if the spatial dimensions of JJs also need to be resolved. Therefore, the ability to bypass the need for high resolutions through coarse-graining makes the method a powerful tool to study multi-scale systems.

We now discuss the coarse-grained equations of motion for superconductors. Consider, for simplicity, a uniform mesh made up of 3D rectangular ``brick" elements. To obtain dynamical equations for the coarse-grained fluxes and charges, we evaluate Eq.\,(\ref{eq:Aprime_waveeq_rho}) at a dual face $e^\dagger$ by performing $\int_{e^\dagger} (\ref{eq:Aprime_waveeq_rho})\cdot{\bf da}$ and evaluate Eq.\,(\ref{eq:chargeconserve2_rho}) at a dual volume $v^\dagger$ by performing $\int_{v^\dagger}(\ref{eq:chargeconserve2_rho}).dV$. The resulting discrete equations are:
\begin{widetext}
%\begin{align}\label{Eq:DiscreteAmpereLaw_full}
%    & \mu\epsilon\frac{\Delta A(e^+)}{\Delta\ell(e)}\ddot{\phi}'(e) + \sum\limits_{e_0\epsilon\partial e^+}\!\sum\limits_{e_1\epsilon\partial e_0^+}\!\frac{\Delta \ell(e_0)}{\Delta A(e_0^+)}\Phi(e_1) + \frac{\Delta A(e^+)}{\lambda_L^2\Delta\ell(e)}\Phi(e) - \frac{\mu q^2}{m}\delta\bar{\rho}(e)\frac{\Delta A(e^+)}{\Delta\ell(e)}\Phi(e) \\
%    - \frac{\mu q}{2m}\frac{\Delta A(e^+)}{\Delta\ell(e)}&\bigg(\epsilon\frac{\partial}{\partial t} + \sigma\bigg)\Delta\bigg(\frac{\Phi^2(e_x)}{\Delta\ell_x^2} + \frac{\Phi^2(e_y)}{\Delta\ell_y^2} + \frac{\Phi^2(e_z)}{\Delta\ell_z^2} \bigg) + \mu\sigma\frac{\Delta A(e^+)}{\Delta\ell(e)}\dot{\Phi}(e) = \frac{\mu q^2}{m}\rho_{src}\frac{\Delta A(e^+)}{\Delta\ell(e)}\Phi(e) + \mu I_{src}(e), \nonumber
%\end{align}
%and
%\begin{equation}\label{Eq:DiscreteChargeEoM}
%    \delta\dot{\rho}(\text{v}) = \frac{1}{\Delta\ell_x\Delta\ell_y\Delta\ell_z}\sum\limits_{e|v\in e}\bigg(\frac{1}{\mu q\lambda_L^2} + \frac{q}{m}\delta\bar{\rho}(e) \bigg)\frac{\Delta A(e^+)}{\Delta\ell(e)}\Phi(e),
%\end{equation}

\begin{align}\label{Eq:DiscreteAmpereLaw_full}
    \ddot{\Phi}(e) + \frac{1}{\mu_0 \epsilon_0}{\dbcurl} \Phi(e) + \frac{1}{\mu_0\epsilon_0\lambda_L^2}\Phi(e) + \frac{q^2}{\epsilon_0 m} \delta\bar{\rho}(e) \Phi(e)- \frac{q}{2m} \frac{\partial}{\partial t}\gradAsq\Phi(e)^2 +& \frac{\mu_0\epsilon_0\hbar^2}{2mq}\frac{\partial}{\partial t}\pressureop\delta\rho(v_{v \subset \partial e}) \nonumber \\
    &= \frac{1}{\epsilon_0}\frac{\Delta\ell(e)}{\Delta A(e^+)} I_{src}(e),
\end{align}
and
\begin{equation}\label{Eq:DiscreteChargeEoM}
    \Delta V(v^{\dagger})\delta\dot{\rho}(v) = \sum\limits_{e|v\in e}\bigg(\frac{1}{\mu q\lambda_L^2} + \frac{q}{m}\delta\bar{\rho}(e) \bigg)\frac{\Delta A(e^+)}{\Delta\ell(e)}\Phi(e),
\end{equation}
respectively, where $\Delta V(v^{\dagger})$ is the volume of the cell dual to the vertex $v$. The curl-curl operator acting on ${\bf A'}$ in Eq.\,(\ref{eq:Aprime_waveeq_rho}) is represented in Eq.\,(\ref{Eq:DiscreteAmpereLaw_full}) by the symbol $\dbcurl$, whose inspiration comes from the geometric construction of the discrete operator acting on the coarse-grained field $\Phi(e)$ (see Fig.\,(\ref{fig:dualmesh_DEC}g)). The symbol $\delta\bar{\rho}$ represents the value of $\delta\rho$ determined along an edge $e$, which is computed by averaging the values of $\delta\rho$ at the end vertices of $e$ (as shown in Fig.\,(\ref{fig:dualmesh_DEC}h)). The field $\bar{\rho}$ is introduced so that the nonlinear term $\rho{\bf A}'$ can be accurately treated in our edge-based discretization scheme~\cite{vectorbundles_dec}. The non-linear term $\nabla\big|{\bf A'}\big|^2$ is represented by the symbol $\gradAsq$ acting on $\Phi(e) \cdot \Phi(e^{\dagger})$ (see Fig.\,(\ref{fig:dualmesh_DEC}k)). Finally, the quantum pressure term $\nabla\bigg[\frac{\nabla^2(\sqrt{\rho})}{\sqrt{\rho}}\bigg]$ is represented by the symbol $\pressureop$ acting on the two vertices at the boundary of $e$. The dashed lines in the symbol indicate that the quantities that go into this term are the scalars at the boundary vertices of the edges, not the edges themselves.

%Similarly, we can also choose to discretize Eqs.\,(\ref{eq:Aprime_waveeq2}) and (\ref{eq:gausslaw2}) so that they become \ks{Same for these as well, would be good to label the terms and cross-reference with Eq 12-13}
%\begin{align}\label{Eq:DiscreteAmpereLaw_full2}
%    0 =& \mu\epsilon\frac{\Delta A(e^+)}{\Delta\ell(e)}\ddot{\phi}'(e) \nonumber\\
%    &+ \sum\limits_{e_0\epsilon\partial e^+}\!\sum\limits_{e_1\epsilon\partial e_0^+}\!\frac{\Delta \ell(e_0)}{\Delta A(e_0^+)}\Phi(e_1) + \frac{\Delta A(e^+)}{\lambda_L^2\Delta\ell(e)}\Phi(e) - \frac{\mu\epsilon q}{m}\bigg(\frac{\Delta\psi(e_x)}{\Delta\ell_x^2} + \frac{\Delta\psi(e_y)}{\Delta\ell_y^2} +\frac{\Delta\psi(e_z)}{\Delta\ell_z^2} \bigg)\frac{\Delta A(e^+)}{\Delta\ell(e)}\Phi(e) \nonumber\\
%    &- \frac{\mu q}{2m}\frac{\Delta A(e^+)}{\Delta\ell(e)}\bigg(\epsilon\frac{\partial}{\partial t} + \sigma\bigg)\Delta\bigg(\frac{\Phi^2(e_x)}{\Delta\ell_x^2} + \frac{\Phi^2(e_y)}{\Delta\ell_y^2} + \frac{\Phi^2(e_z)}{\Delta\ell_z^2} \bigg) + \mu\sigma\frac{\Delta A(e^+)}{\Delta\ell(e)}\dot{\Phi}(e) - \frac{\mu q^2}{m}\rho_{src}\frac{\Delta A(e^+)}{\Delta\ell(e)}\Phi(e) \nonumber\\
%    &  -\mu I_{src}(e),
%\end{align}
%and
%\begin{align}
%    \sum\limits_{e|v\in e}\frac{\Delta A(e^+)}{\Delta \ell(e)}\dot{\psi}(e^+) =  \sum\limits_{e|v\in e}\Bigg[ -\!\frac{1}{\epsilon\mu \lambda_L^2} + \frac{q}{m}\bigg(\frac{\Delta\psi(e_x)}{\Delta\ell_x^2} + \frac{\Delta\psi(e_y)}{\Delta\ell_y^2} +\frac{\Delta\psi(e_z)}{\Delta\ell_z^2} \bigg)\Bigg]\frac{\Delta A(e^+)}{\Delta\ell(e)}\Phi(e) - \frac{\dot{Q}_{src}}{\epsilon},
%\end{align}
%respectively.
\end{widetext}
Eqs.\,(\ref{Eq:DiscreteAmpereLaw_full}) and (\ref{Eq:DiscreteChargeEoM}) allow us to time-evolve the fields $(\Phi(e), \delta\rho(v))$ at every edge $e$ and every vertex $v$ in the computational domain, respectively. The discretization of time can be done in a similar manner as in the standard FDTD method, in which the time derivative term $\ddot{\Phi}(e)$ in Eq.\,(\ref{Eq:DiscreteAmpereLaw_full}) evaluated at the $n^{\text{th}}$ time step is given by
\begin{equation}\label{eq:discrete_time2nd_derivative}
    \ddot{\Phi}(e)|_{t_n} = \frac{\Phi(e)_{t_{n+1}}-2\Phi(e)|_{t_n}+\Phi(e)|_{t_{n-1}}}{\Delta t},
\end{equation}
where $\Delta t$ is the uniform time step, and $t_n=n\Delta t$. Inserting the discrete form for $\ddot{\Phi}(e)$ given in Eq.\,(\ref{eq:discrete_time2nd_derivative}) into Eq.\,(\ref{Eq:DiscreteAmpereLaw_full}), the value of the flux field at the next time step, $\Phi(e)|_{t_{n+1}}$, can be found in terms of flux fields and charges at the current time step. The propagation in time of $\delta\rho(v)$ according to Eq.\,(\ref{Eq:DiscreteChargeEoM}) follows a similar strategy. 

Note that in Eqs.\,(\ref{Eq:DiscreteAmpereLaw_full}) and (\ref{Eq:DiscreteChargeEoM}), we have split the charge density $\rho$ into the background part $\rho_0$ and the fluctuation $\delta\rho$, but still keep all the nonlinear terms as in the full Eqs.\,(\ref{eq:Aprime_waveeq_rho}) and (\ref{eq:chargeconserve2_rho}). Each of the nonlinear terms in Eq.~(\ref{Eq:DiscreteAmpereLaw_full}) are treated differently in DEC. The $\delta\rho{\bf A}'$ term is a product of two entities occupying different regions in space, i.e. a scalar living on primal vertices ($\delta\rho$) with a vector field defined on edges (${\bf A}'$). Therefore the edge variable $\bar{\rho}$ is introduced as a natural solution to perform the product between a scalar and a vector. The nonlinearity in the $\nabla\big|{\bf A'}\big|^2$ term, on the other hand, comes from the dot product ${\bf A}'.{\bf A}'$ of two vectorial quantities that supposedly both live on edges, but the product needs to produce a scalar that live on vertices. One therefore need to invoke the definition of a support volume. An example of a support volume of a primal edge is shown in Fig.\,(\ref{fig:dualmesh_DEC}i), while a formal definition is given in Def 2.4.9 of Ref~\cite{DEC_HiraniThesis}. The dot product defined on a support volume $V_{s}(e)$ of an edge $e$ is then written as
\begin{align}
    \int_{V_s}|{\bf A}'|^2dV &= \Phi(e) \cdot \Phi(e^{\dagger}) \nonumber \\
    &= \frac{\Delta A(e^\dagger)}{\Delta\ell(e)}\Phi(e)^2.
\end{align}
To find the dot product on each vertex (or equivalently, on the dual volume) we then compute the sum over all the edges attached to that vertex as follows
\begin{equation}
    |{\bf A}'|^2(v) = \frac{1}{\Delta V(v^{\dagger})}\sum_{e\supset v} \frac{\Delta V(v^{\dagger})\cap V_{s}(e)}{\Delta V(v^{\dagger})}\frac{\Delta A(e^\dagger)}{\Delta\ell(e)}\Phi(e)^2.
\end{equation}
The final form of $\nabla\big|{\bf A'}\big|^2$ is therefore defined on each of the primal edges $e$ 
\begin{equation}
    \int_e \nabla\big|{\bf A'}\big|^2{\bf d\ell} = |{\bf A}'|^2(v_B) - |{\bf A}'|^2(v_A),
\end{equation}
where $v_A$ and $v_B$ are the boundary vertices of $e$. This operation uses values of the field on the edges surrounding $v_A$ and $v_B$, as illustrated by the operator $\gradAsq$. Using Eqs.\,(\ref{Eq:DiscreteAmpereLaw_full}) and (\ref{Eq:DiscreteChargeEoM}), it is possible to analyze the dynamics of both material and electrodynamic degrees of freedom in any system of superconducting structures with controllable accuracy. Finally, the treatment of the quantum pressure term is rather straightforward, as one only need to apply a discrete Laplacian followed by a discrete gradient operator. The bullets in the symbol $\pressureop$ indicate the vertices involved in the action of this operator. In Eq.\,(\ref{Eq:DiscreteAmpereLaw_full}) this operator acts on $\delta\rho$ instead of $\rho$, as we have made the assumption that $\rho_0$ is static and does not contribute to the variation in time of the pressure term. Note that we can obtain the discrete version of Eq.\,(\ref{eq:Aprime_waveeq}) simply by discarding the discrete pressure term from Eq.\,(\ref{Eq:DiscreteAmpereLaw_full}). The graphical notations that we introduced in Eq.\,(\ref{Eq:DiscreteAmpereLaw_full}) above serve both as geometrical visualizations and short-hands for the more lengthy mathematical representations of the operators involved. 
For completeness, in Appendix \ref{append:discreteMaxwell} we also derive the full form of these operators.

This full 3+1D flux-based formulation can be used to reproduce the results of the 1+1D theory of a transmission line~\cite{1+1D_Moein_2016, blaisrmp} through an appropriate limiting procedure, as demonstrated in Appendix \ref{append:1Dderivations}. The equation of motion for the flux field in a 1D waveguide whose longitudinal dimension is along $x$ is given by
\begin{equation}\label{eq:wave_eq_1D_Psiz}
    \partial_x^2\Psi_z - lc \, \partial_t^2\Psi_z = 0,
\end{equation}
where $\Psi_z(x) = \int\!\psi(e_z(x))dt$ is the flux along the edge $e_z$ that spans $z$, one of the transverse dimensions of the waveguide. The capacitance $c$ and inductance $l$ per unit length depends on the material properties and geometrical dimensions of the waveguide. 
Notice that $\Psi_z^i$ in Eq.\,(\ref{eq:wave_eq_1D_Psiz}) is indeed the usual flux variable $\int\!dt \, V$ used in lumped-element treatment of cQED. Moreover, one can show that the flux through a unit cell in the 1D discretization of the waveguide is the same as the difference of $\Psi$ defined at the two nodes of a cellular inductor in the lumped-element circuit. Therefore, from the full 3D perspective we can have a correct physical interpretation for the lumped-element-based circuit theory of a one-dimensional transmission line. Detailed derivation of Eq.\,(\ref{eq:wave_eq_1D_Psiz}) from 3+1D formulation and discussions on the equivalence between the 3D Maxwell formulation and the 1D transmission line circuit theory is presented in Appendix \ref{append:1Dderivations}.
%  \ks{this is awesome!!!}

The focus of this work is on the accurate numerical simulation of 2D and 3D systems. In the subsequent sections, we will provide several examples to demonstrate the ability of the coarse-grained computational model to accurately capture the known physics of superconducting materials interacting with an electromagnetic environment.

\section{Numerical results}\label{sec:numerics}
\subsection{Linear modes of the system}\label{sec:linearmodes}
%\begin{figure*}[t]
%    \centering
%    \includegraphics[scale=0.15]{PhiX_Mode1.png}
%    \includegraphics[scale=0.15]{PhiX_Mode2.png}
%    \includegraphics[scale=0.15]{PhiX_Mode3.png}
%    \includegraphics[scale=0.15]{PhiX_Mode4.png}
%    \includegraphics[scale=0.15]{PhiX_Mode5.png}
%    \caption{The first five modes of $a_{1x}$ satisfying Eq.\,(\ref{eq:homogenNorm_a1}) are shown. The system studied is a 20-by-20 cavity surrounded by superconducting walls with $\lambda_L=0.1$.}
%    \label{fig:eigenmodes_Afield}
%\end{figure*}
\begin{figure*}[t]
    \centering
    \includegraphics[scale=0.21]{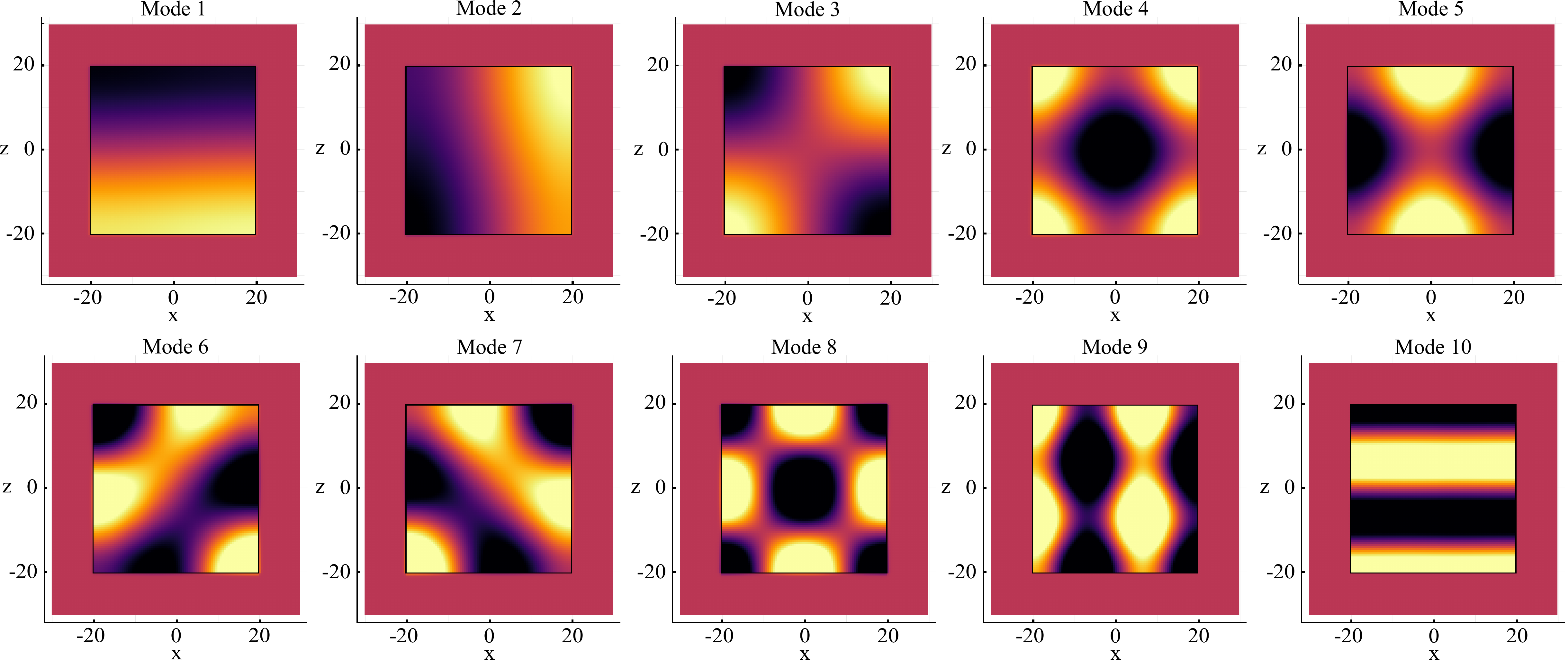}
    \caption{The flux fields of first ten modes, given by $\Phi_y = \int{\bf B}_{1}.{\bf dS}_{xz}=\int({\bf \nabla}\times{\bf a}_1).{\bf dS}_{xz}$, with ${\bf a}_1$ satisfying Eq.\,(\ref{eq:linearized_ampere}) are shown. The system studied is a 40-by-40 cavity, in units of $\lambda_L$, surrounded by superconducting walls. The entire computational domain is subdivided into 2D rectangular bricks, and the resulting flux fields $\Phi_y$ plotted here are the fluxes that thread through the individual rectangular pixels.}
    \label{fig:eigenmodes_Bfield}
\end{figure*}

The computational model discussed in Sec.\,\ref{sec:discrete_EM_formulation} can be readily applied to the calculation of linear modes of the system. The linear part of Eq.\,(\ref{eq:Aprime_waveeq}) in steady-state is given by
\begin{equation}\label{eq:linearized_ampere}
    {\bf \nabla}\!\times\!{\bf \nabla}\!\times\!{\bf a}_1 -  \bigg(\mu_0\epsilon_0 \omega^2 -\frac{1}{\lambda_L^2}\bigg){\bf a}_1 = 0,
\end{equation}
where ${\bf a}_1$ is the spatial component of the linearized field ${\bf A}'={\bf a}_1e^{-i\omega t} + {\bf a}^*_1e^{i\omega t}$ (see Appendix \ref{append:perturbationtheory}). 
The eigenmodes of Eq.\,(\ref{eq:linearized_ampere}), can be  computed using standard eigensolvers after the discretization described in Sec.\,(\ref{sec:discrete_EM_formulation}) is performed. To demonstrate this, we calculate the eigenmodes of fields trapped in a square cavity with superconducting boundaries, as shown in Fig.\,(\ref{fig:eigenmodes_Bfield}). We assume translational invariance along $y$ so that the cavity is effectively 2D. The superconductor surrounding the cavity has a characteristic penetration depth $\lambda_L$. %%%he field $a_{1x}$ of the first 8 modes are shown in Fig.\,(\ref{fig:eigenmodes_Afield}).
The resulting flux field $\Phi_y$ normal to the 2D surface, given by $\Phi_y = \int{\bf B}_1.{\bf dS}_{xz} = {\int(\bf \nabla}\times{\bf a}_1).{\bf dS}_{xz}$ are shown in Fig.\,(\ref{fig:eigenmodes_Bfield}). In the limit $\lambda_L\rightarrow 0$, the eigenspectrum is discrete due to the finite confinement of the field and is given by
\begin{equation}\label{eq:eig_squarecav}
    E_{m,n} = \mu_0\epsilon_0\omega^2 = \frac{\pi^2}{L^2}(m^2 + n^2),
\end{equation}
where $L$ is the size of the cavity. In Table\,\ref{table:eig_lambdaL} we present the eigenvalues of the first five modes when different values of the ratio $\Tilde{\lambda}=\lambda_L/L$ is considered. The eigenvalues converge to the analytical values in Eq.\,(\ref{eq:eig_squarecav}) when $\lambda_L$ decreases. The numerically calculated values for when $\lambda_L=0$ agrees well with the analytical values.

\begin{table}[ht]
	\centering
	\renewcommand{\arraystretch}{1.5}
	\begin{tabular}{c|c|c|c|c|c}
          \toprule
          Mode & $\Tilde{\lambda}\!=0.1$ & $\Tilde{\lambda}\!=0.03$ & $\Tilde{\lambda}\!=0.01$ & $\Tilde{\lambda}\!= 0$ & $\Tilde{\lambda}\!\!\rightarrow\!0$ (analytical)\\
          \hline
          $1^{\text{st}}$ & 1.1578 & 1.0307 & 1.0055 & 1.0012 & 1.0 \\
          \hline
          $2^{\text{nd}}$ & 1.1441 & 1.0311 & 1.0055 & 1.0014 & 1.0 \\
          \hline
          $3^{\text{rd}}$ & 1.6789 & 1.4714 & 1.4242 & 1.4135 & $\sqrt{2}$ \\
          \hline
          $4^{\text{th}}$ & 2.2713 & 2.0603 & 2.0106 & 2.0018 & 2.0 \\
          \hline
          $5^{\text{th}}$ & 2.2719 & 2.0608 & 2.0112 & 2.0021 & 2.0 \\
 \botrule
 \end{tabular}
 \caption{The values of $L\sqrt{E_{m,n}}/\pi$ for the first five  modes of the ${\bf a}_1$ field satisfying Eq.\,(\ref{eq:linearized_ampere}) are shown. The system studied is a cavity surrounded by superconducting walls.}
	\label{table:eig_lambdaL}
\end{table}

\subsection{Dipole source in a superconducting cavity}\label{sec:oscillating_dipole}
\begin{figure}[t]
    \centering
    \includegraphics[scale=0.25]{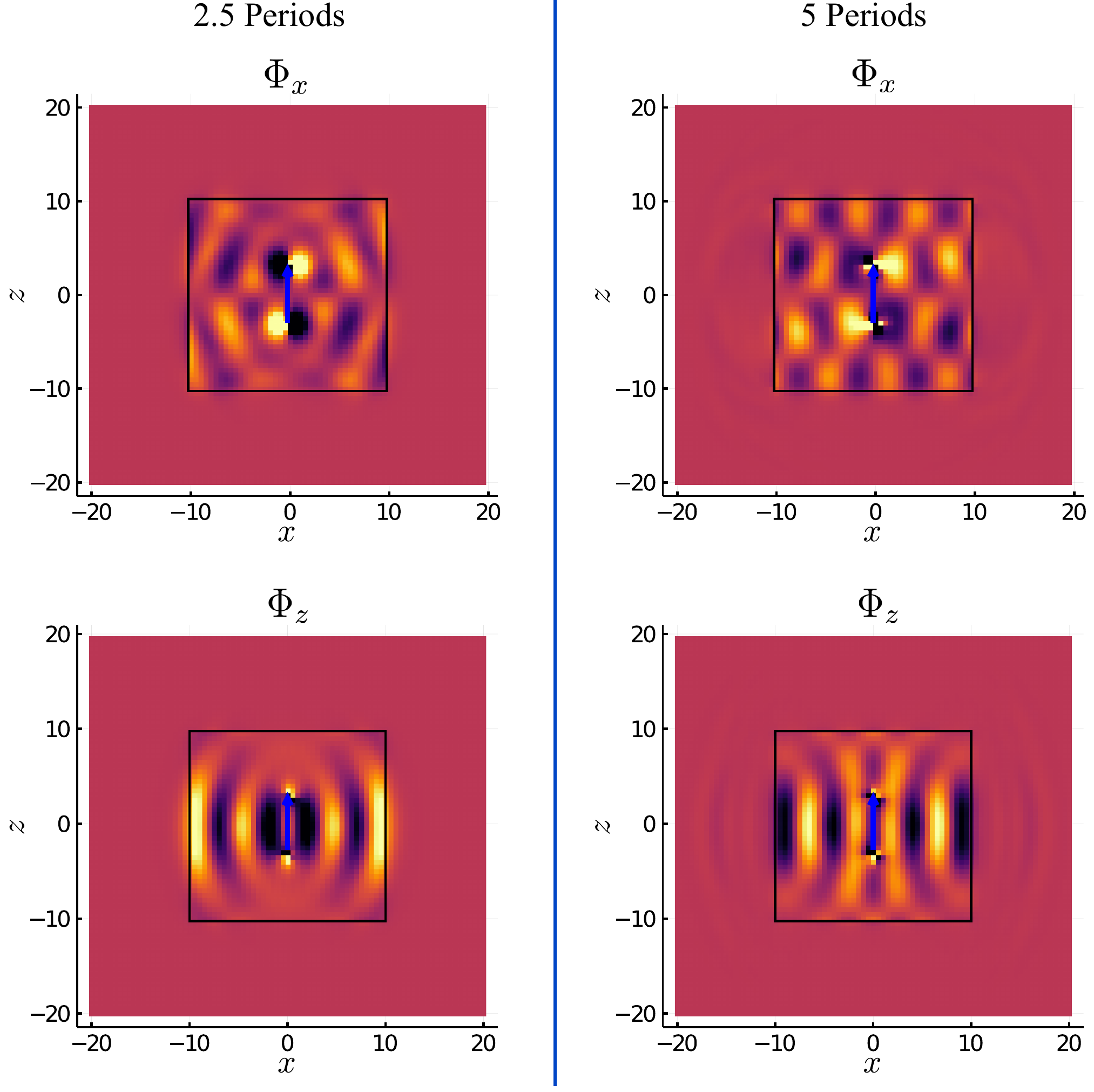}
    \caption{The $\Phi_x$ and $\Phi_z$ fields of a dipole inside a cavity surrounded by superconducting boundaries. The dipole, denoted by the blue arrow, oscillates in a wavelength of $\lambda=5$, and the cavity size is $20\times 20$. All lengths are in units of $\lambda_L$. The column on the left shows the fields after 2.5 periods, while the column on the right presents the fields after 5 periods.}
    \label{fig:dipole_cavity}
\end{figure}

The formulation presented in Secs.\,\ref{sec:formulation} and \ref{sec:discrete_EM_formulation} is specifically designed for accurately capturing the dynamics of interactions between the superconducting condensate field and electromagnetic fields. As an initial example, we consider the case of an oscillating dipole source within a rectangular cavity bounded by superconducting walls. In the discretized geometry, the dipole is modeled by a current flowing along a vertical line that is formed by connected edges and is terminated by two vertices carrying the dipole charges \mbox{$\pm Q(t) = \pm Q_0(1-\cos(\omega t))$}. This choice of time-dependent dipole charge ensures that at $t=0$ the system is completely unexcited and that conservation of sourced charges is guaranteed at all times. The time dynamics of this dipole-cavity system is shown in Fig.\,(\ref{fig:dipole_cavity}), where a snapshot of the coarse-grained fields $\Phi_x$ and $\Phi_z$  are plotted at two different times. The notations $\Phi_x$ and $\Phi_z$ denote the edge fields defined in Eq.~(\ref{eq:phip_define}), now evaluated at the $x-$oriented edges $e_x$ and $z-$oriented edges $e_z$ of the square lattice, respectively. The size of the cavity studied is $20\times 20$, and the oscillations excited by the dipole has a wavelength of $\lambda=5$, all in units of the penetration depth $\lambda_L$ of the superconducting boundary. From Fig.\,(\ref{fig:dipole_cavity}) we can see that at $t = 2.5T$, where $T = \lambda/c$ is the period of oscillation, the field has just reached the boundaries of the cavity. At this point we start to see the wave being reflected from the walls, with a small leakage into the material. The use of hybridized generalized flux field $\Phi$ in our formulation allows for a straightforward representation of the light-matter dynamics both in the vacuum region and inside the material. At $t = 5T$, when the reflected waves have reached the dipole source, the electromagnetic pattern in the cavity is the result of interference of the waves generated by the dipole and the reflected waves from the cavity walls. Because there is no damping mechanism these oscillations will reverberate indefinitely.  

\subsection{The Meissner effect}\label{sec:meissner_effect}
\begin{figure}[H]
    \centering
    \includegraphics[scale=0.4]{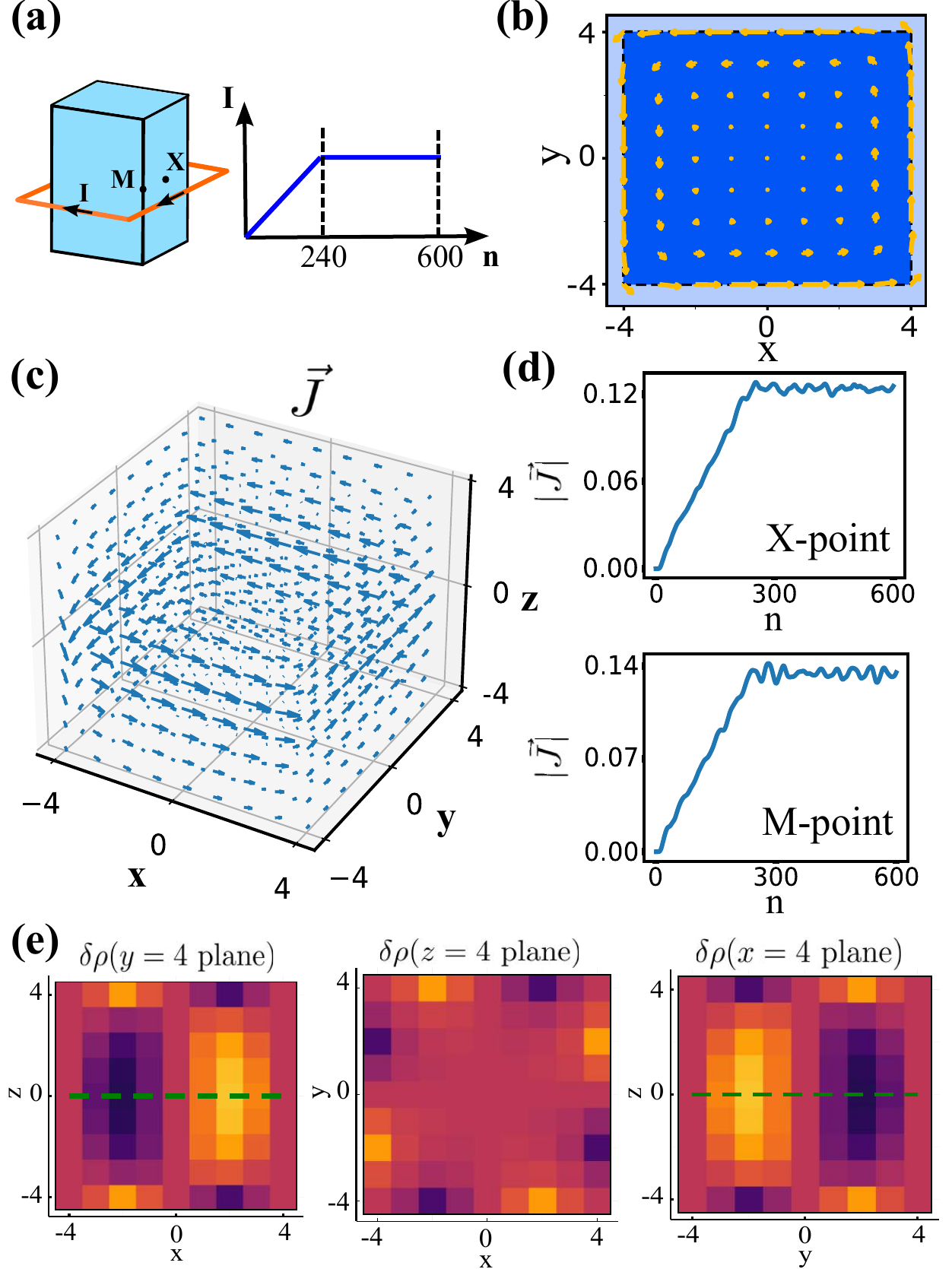}
    \caption{(a) A schematic of the system studied: we consider a superconducting cuboid under the influence of external time-varying magnetic field created by a current loop wrapped around the superconducting piece. The dependence in time of the current in the external loop is also shown. (b) The current density ${\bf J}$ inside the superconducting cuboid at the $z=0$ plane and at time step $n=300$. (c) The current density inside the superconducting cuboid at time step $n=300$. The cuboid is $8\!\times\! 8\!\times\!8$ in unit of $\lambda_L$. (d) The amplitude $|{\bf J}|$ over time at the symmetry points (X-point and M-point) of the cuboid. (e) The charge fluctuation $\delta\rho$ on the surface of the superconducting piece at $n=300$: from left to right, the $xz-$plane at $y=4$, $xy$-plane at $z=4$, and $yz-$plane at $x=4$ are shown. The green dashed line corresponds to the external current loop viewed from the side.}
    \label{fig:meissner_results}
\end{figure}

The formulation presented in this paper enables the simulation of the time-dependent response of a superconductor to an external magnetic field and the observation of the transient stage of this magnetic repulsion. To demonstrate this, in Fig.\,(\ref{fig:meissner_results}), we show the time-domain results of simulating the dynamics of fields by solving Eq.\,(\ref{Eq:DiscreteAmpereLaw_full}) and (\ref{Eq:DiscreteChargeEoM}). The system of interest is a piece of three-dimensional superconducting material interacting with an external magnetic field created by a current loop (Fig.\,(\ref{fig:meissner_results}a)). The current in the loop exhibits a linear ramp behavior over time initially, after which it remains constant until the end of the simulation. As shown in Figs.\,(\ref{fig:meissner_results}b) and (\ref{fig:meissner_results}c), currents are generated inside the superconductor near the surface to nullify the penetration of the field created by the external loop. 
The hybridized field ${\bf A}'$, along with the internal current ${\bf J}$ (given by ${\bf J}\sim\rho{\bf A}'$ as shown in Eq.~(\ref{eq:supercurrent_def})), decays towards the bulk of the material, which is consistent with the Meissner effect~\cite{MeissnerPaper}.

As the current in the external loop enters the steady state, so does the dynamics inside the superconductor. The dependence in time of current densities at symmetry points of the superconductor are shown in Fig.\,(\ref{fig:meissner_results}d), where we see the currents fluctuate around the steady state values. These oscillations will reverberate indefinitely because of the lack of any dissipation mechanism. The generation of a non-uniform supercurrent also brings about redistribution of charges in the superconductor, particularly near the surface. In Fig.\,(\ref{fig:meissner_results}e), we show the charge distribution on the surface of the superconductor. Due to the highly symmetric object considered, there are symmetry points at which the charge distribution remains neutral. This is most evidently seen in the middle plot in Fig.\,(\ref{fig:meissner_results}e), where the top view of the upper surface of the superconductor is shown. We see that the charge is neutral along the diagonals and along the lines connecting the midpoints of opposite boundaries -- all are symmetry lines of a square. These oscillations are indicative of a charge-density wave formation near the surface of the superconductor.

\subsection{Flux quantization}\label{sec:flux_quantization}
\begin{figure}[H]
    \centering
    \includegraphics[scale=0.5]{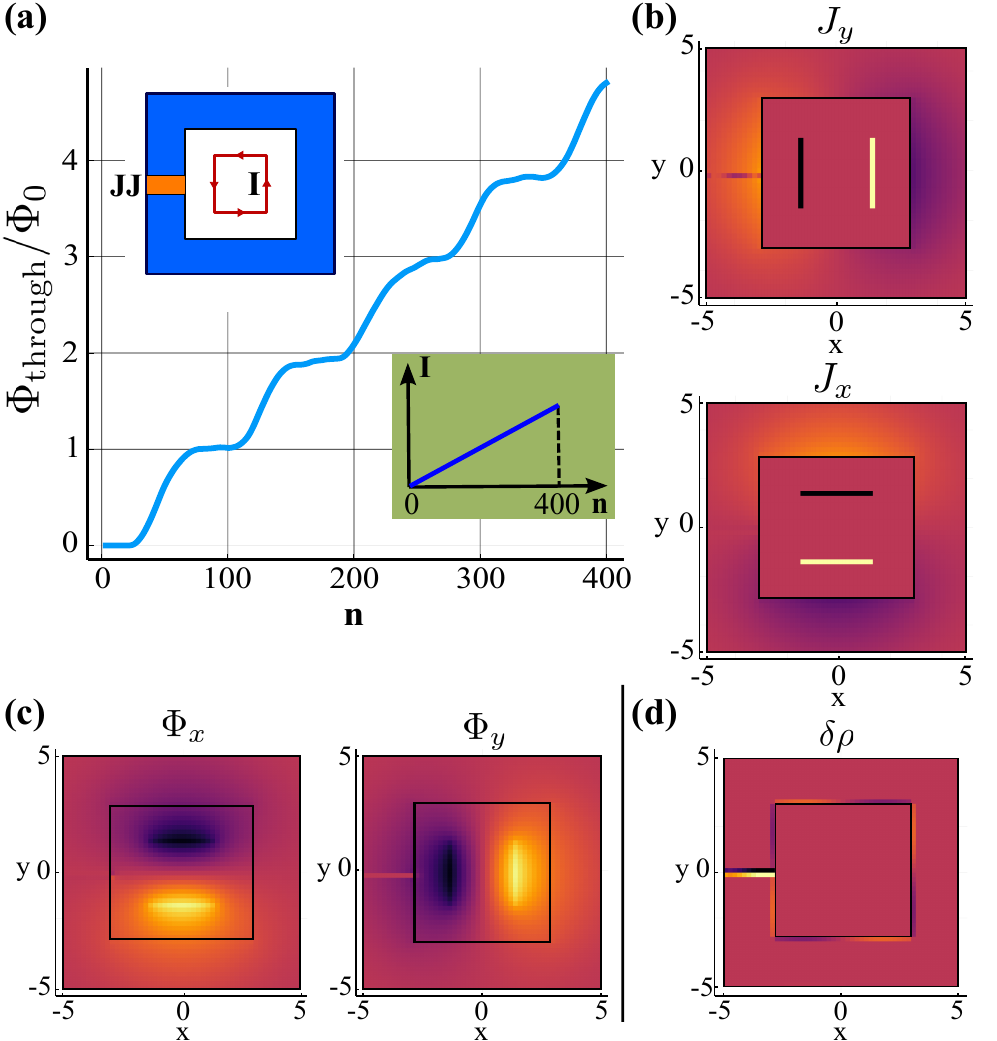}
    \caption{(a) The dependence in time of calculated fluxoid (in flux quantum unit) trapped inside the superconducting loop. The top-left inset depicts the 2D geometry studied: an infinitely long superconducting tube that has a Josephson slit. The tube is $10\!\times\!10$ wide, with a thickness of $2$, all in units of $\lambda_L$. There is also a coil inside the tube carrying a controlled time-varying current. The bottom-right inset shows the time-dependence of the controlled current. (b) The current densities $J_y$ (upper plot) and $J_x$ (lower plot) on the 2D plane at time step $n=60$. (c) The $\Phi_x$ and $\Phi_y$ fields at $n=60$. (d) Charge distribution $\delta\rho$ on a 2D plane at $n=60$.}
    \label{fig:2d_fluxquantization}
\end{figure}

Flux quantization is another fundamental feature of superconductors that can be observed in macroscopic devices~\cite{FluxQuantization1_1961, FluxQuantization2_1961}. In this section, we will examine the time-dependent behavior of the magnetic flux trapped in a non-simply connected superconducting object as it transitions between quantized values under the influence of a smoothly varying external magnetic field.
\begin{figure}[H]
    \centering
    \includegraphics[scale=0.38]{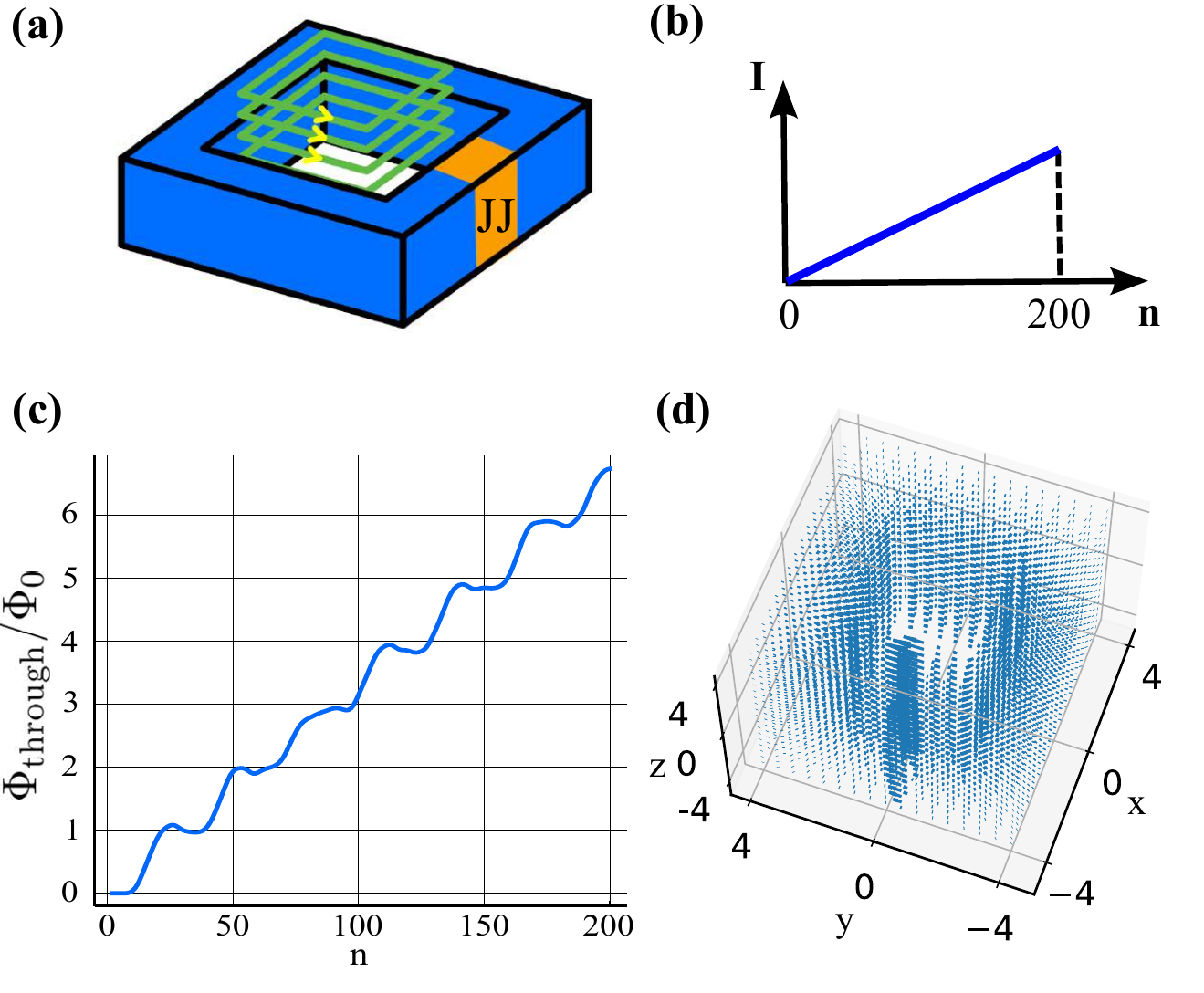}
    \caption{(a) A schematic of the system studied: we consider a superconducting loop with a finite thickness and a concentric coil of finite length placed inside the hole. The superconducting piece is $8\!\times\!8\times\!8$ in size, while the current loop is $2\!\times\!2\!\times\!8$, all in units of $\lambda_L$. (b) The dependence in time of the current in the coil. (c) The calculated fluxoid (in flux quantum unit) trapped inside the superconducting loop as a function time step. (d) The density of supercurrent inside the superconductor at time step $n=96$.}
    \label{fig:3d_fluxquantization}
\end{figure}
The first object we consider is an infinitely long, hollow, concentric superconducting tube with a square cross section and a finite thickness. On the periphery of the tube there is a slit where an insulator is placed to form a Josephson junction (see the inset in Fig.\,(\ref{fig:2d_fluxquantization}a)).
The primary role of the junction here is to serve as an insulating region where quasiparticles (normal electrons) are allowed to exist. These normal electrons can then give rise to vortices that can travel within this insulating channel. Due to the infinite length of the tube, the fields are translationally invariant along that dimension and the problem is effectively 2D. The quantization of a fluxoid~\cite{London_superfuilds} in this case is given by
\begin{equation}\label{eq:fluxoid_def}
N\Phi_0 = \oint {\bf A} \cdot {\bf d\ell} + \Phi_j ,
\end{equation}
where the closed integral is performed in a loop inside the interior of the superconducting tube, $N$ is the number of flux quanta in the loop, $\Phi_0$ is a flux quantum, and $\Phi_j$ is the value of $\Phi$ across the junction. Inside the tube a coil carrying a current is placed. The current is controlled to increase linearly over time from zero. The simulation domain is truncated at a finite distance away from the system of interest, where the Dirichlet boundary condition for perfect a superconductor (i.e. $A_t$, the tangential component of ${\bf A}'$, vanishes) is imposed. The results of our time-domain calculation are presented in Fig.\,(\ref{fig:2d_fluxquantization}). As shown in Fig.\,(\ref{fig:2d_fluxquantization}a), when the current increases, the flux trapped inside the tube follows a step-wise behavior, with the steps residing approximately at integer multiples of flux quanta. The value of $\Phi_j$, which is needed to determine the trapped flux according to Eq.~(\ref{eq:fluxoid_def}), is computed by summing up the values of $\Phi$ (defined in Eq.~(\ref{eq:phip_define})) on the edges along a traversal line that connects the two junction-superconductor interfaces.

In Figs.\,(\ref{fig:2d_fluxquantization}b) and (\ref{fig:2d_fluxquantization}c) we plot the density of supercurrent and the $\Phi$ field, respectively, inside the superconducting tube. Within and near the junction there are currents both in transversal and longitudinal directions with respect to the junction. The transversal current shields the superconducting bulk from the external field created by the coil, while the longitudinal current is needed to transport the vortices with normal electron cores between the inner and outer boundaries of the tube to enable jumps in the quantized flux. The spatially non-uniform dynamics near the junction also creates charge imbalances, which are shown in Fig.\,(\ref{fig:2d_fluxquantization}d).

Next we consider a variant of the previous geometry which can not be reduced to a two-dimensional domain. The set up is similar to that of the infinite tube case, with the only difference being the lengths of the tube and the coil are now finite (see Fig.\,(\ref{fig:3d_fluxquantization}a)), which makes the problem fully 3D. Again we observe the approximate quantization of flux in the SQUID loop, as can be seen in Fig.\,(\ref{fig:3d_fluxquantization}c). The current density is shown in Fig.\,(\ref{fig:3d_fluxquantization}d), where a highly non-uniform spatio-temporal dynamics can be seen near the junction region. The DEC equations are able capture the boundary-layer dynamics that is otherwise difficult to capture. 

\begin{figure}[H]
    \centering
    \includegraphics[scale=0.8]{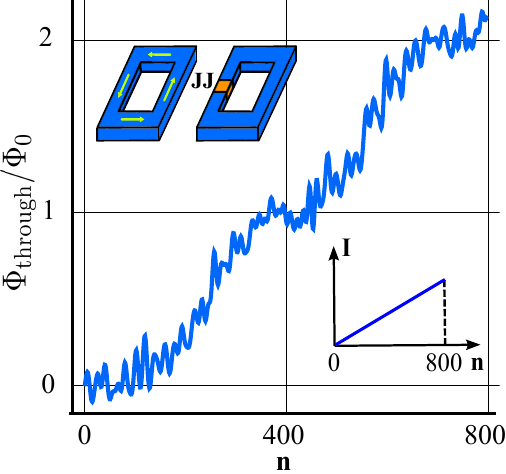}
    \caption{The calculated fluxoid (in flux quantum unit) trapped inside the superconducting loop over time. The top-left inset describes the system studied: a superconducting loop that has a narrow JJ slit. A second loop carrying a controlled current is placed next to the primary loop. The dimensions of both loops in units of $\lambda_L$ are $12\!\times\!12\!\times\!4$, and the thickness of the superconducting loop is $4$. The linear dependence in time of the controlled current is shown in the bottom right inset.}
    \label{fig:3d_fluxquanta_JJsidebyside}
\end{figure}

We now turn to a more realistic geometry where flux quantization is relevant. This is the case when the current source is placed outside and next to the superconducting loop (the inset in Fig.\,(\ref{fig:3d_fluxquanta_JJsidebyside})). The current in the loop creates magnetic flux lines that would then enter the superconducting loop from above through fringing fields and magnetically bias the loop containing a weak junction. This geometry is relevant for flux-biasing frequency-tunable Josephson junction qubits \cite{transmonpaper, fluxoniumpaper}. We simulate a situation where the bias current is adiabatically ramped up in a current loop adjacent to the qubit. A more realistic scenario would involve an open current loop, which we will consider in future work. The resulting flux in the SQUID over time is shown in Fig.\,(\ref{fig:3d_fluxquanta_JJsidebyside}), where we see a step-wise quantization, accompanied by noise. The noise comes from various sources that are not present in the previous case with a concentric system, one of which is the interference due to flux lines that reflect from the top and bottom boundaries of the computational domain before entering the superconducting ring. Another source of noise is field lines that penetrate the superconducting ring from the side near the coil, while the JJ is also probed by the external EM field from various directions. Due to part of the noise coming from interference of fields reflected from various boundaries, we suspect the noise will be mitigated if the computational domain is extended further away from the coils and if a more refined mesh is used.

\subsection{Dynamics of the Josephson junction}\label{sec:JJdynamics}
\subsubsection{The Josephson current-phase relation}
The presence of the Josephson junction in our simulations so far in this paper has been done through a direct imposition of the Josephson current-phase relation on the space occupied by the junction. However, as discussed before, our adoption of the coarse-grained hybridized field $\Phi(e) = \int_e{\bf d\ell} \cdot {\bf A}'$, ${\bf A}' = {\bf A} - \frac{\hbar}{q}\nabla\theta$ as the fundamental field to express the equations governing the electrodynamics of superconductors was motivated, in part, by its connection to the flux variable $\varphi$ in the definition of the Josephson phase across a junction. Therefore it should be possible to capture the Josephson effect from the ab-initio equations Eqs.\,(\ref{eq:Aprime_waveeq}) and (\ref{eq:chargeconserve2}) of the hybridized field, or their discrete versions, Eqs.\,(\ref{Eq:DiscreteAmpereLaw_full})-(\ref{Eq:DiscreteChargeEoM}).  In this section, we will demonstrate by way of analytical derivation and numerical simulations how our equations contain the standard physics of a Josephson junction and sub-leading corrections to it.  

\begin{figure}[H]
    \centering
    \includegraphics[scale=0.4]{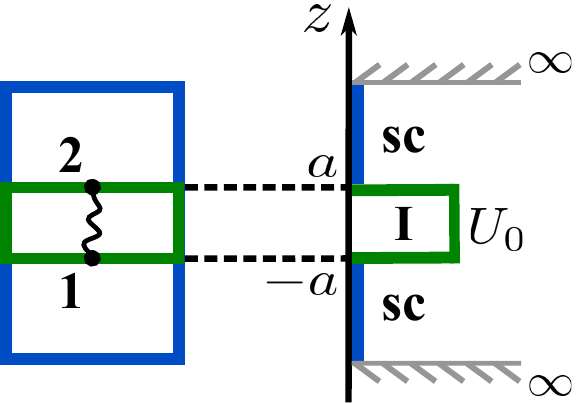}
    \caption{A schematic of a Josephson junction with a 1D potential profile.}
    \label{fig:JJschematic}
\end{figure}

Consider the first Josephson equation~\cite{JosephsonReview_1965}, which states that the current flowing through a JJ is related to the gauge-invariant phase $\varphi$ across it through
\begin{equation}\label{eq:JJ_current}
    J = J_c\sin\varphi,
\end{equation}
where $J_c$ is the critical current density in the junction.  A derivation of this relationship is provided below.

A JJ can be effectively modeled as an insulator sandwiched between two superconducting electrodes, as schematically shown in Fig.\,(\ref{fig:JJschematic}). The one-dimensional potential profile of a bare JJ with a longitudinal length $2a$ is then given by
\begin{equation}
U(z) =  \begin{cases}
               U_0, \hspace{0.07in}|z|<a\\
               0, \hspace{0.15in} |z|>a.
            \end{cases} 
\end{equation}
We assume that the current is directed along the $z$-direction and is uniform in space throughout the insulating region of the junction. Therefore \mbox{${\bf A}' \!=\! A'(z)\hat{z}$} and \mbox{${\bf \nabla}\!\times\!{\bf \nabla}\!\times\!{\bf A}'=0$}, and Eq.\,(\ref{eq:Aprime_waveeq_rho}) applied to the insulating region of the JJ reduces to
\begin{align}\label{eq:1d_A'eq}
    \epsilon_0\ddot{{\bf A}}' + q{\bf J_s} -\frac{\epsilon_0 q}{2m}\frac{\partial}{\partial t}\nabla\big|{\bf A'}\big|^2 + \frac{\epsilon_0\hbar^2}{2mq}\frac{\partial}{\partial t}\nabla\bigg[\frac{\nabla^2(\sqrt{\rho})}{\sqrt{\rho}}\bigg] = 0.
\end{align}
In our model, the supercurrent uniformly flows from one side of the junction to the other, and builds up charges at $\pm a$, where it contacts the superconducting electrodes. Therefore, equivalently, one can also write Eq.\,(\ref{eq:1d_A'eq}) as follows
\begin{align}\label{eq:1d_A'eq2}
    \epsilon_0\ddot{A}' + \frac{\partial^2}{\partial t\partial z}\bigg[(z+a)Q(t)& - \frac{\epsilon_0 q}{2m}\big|A'\big|^2 \nonumber\\ + &\frac{\epsilon_0\hbar^2}{2mq}\frac{1}{\sqrt{\rho}}\frac{\partial^2(\sqrt{\rho})}{\partial z^2} \bigg] = 0,
\end{align}
where $Q(t)$ is the charge built up at the boundaries of the insulating region. To continue, we make the following assumptions: first, we assume that fields vary adiabatically. In other words, we work in the regime where the fields vary at a rate slow enough for the assumption on adiabaticity is satisfied. The first term in Eq.\,(\ref{eq:1d_A'eq2}) can therefore be neglected. We also assume that the current is weak enough so that the the charge built up at the interfaces does not influence the distribution of ${\bf A}'$ and $\rho$ in the insulator. This is equivalent to considering a very thin insulator such that the change in electric potential is negligible throughout its longitudinal dimension. With this assumption, the first term in the bracket in Eq.\,(\ref{eq:1d_A'eq2}) can be neglected too. For initial condition, we assume at $t=0$ the junction is neutral everywhere, and combining with the assumptions already made, we arrive at
\begin{equation}\label{eq:1d_JJeq}
    \big|A'\big|^2 - \frac{\hbar^2}{q^2}\frac{1}{\sqrt{\rho}}\frac{\partial^2(\sqrt{\rho})}{\partial z^2} = 0.
\end{equation}
From Eq.\,(\ref{eq:1d_JJeq}) and the definition of supercurrent in Eq.\,(\ref{eq:supercurrent_def}) we obtain
\begin{equation}\label{eq:1d_JJeq_rho}
    y'' - \frac{1}{y^3} = 0,
\end{equation}
where \mbox{$y = \sqrt{\rho}/\alpha$}, with \mbox{$\alpha = \sqrt{mJ/\hbar |q|}$}. Solving Eq.\,(\ref{eq:1d_JJeq_rho}) with boundary conditions \mbox{$\rho(-a)=\rho_1$} and \mbox{$\rho(a)=\rho_2$} we obtain

\begin{widetext}
\begin{equation}\label{eq:rho_JJinsulator}
    \rho(z) = \bigg[\frac{(\rho_1+\rho_2) - 2\sqrt{-4a^2\alpha^4 + \rho_1\rho_2}}{4a^2} \bigg]z^2 + \frac{(\rho_2-\rho_1)}{2a}z + \frac{\rho_1+\rho_2+2\sqrt{\rho_1\rho_2 - 4a^2\alpha^4}}{4}.
\end{equation}
\end{widetext}
The Josephson phase is then given by
\begin{align}\label{eq:JJ_phase_int}
    \varphi = -\frac{2\pi}{\Phi_0}\int{\bf A}'\cdot {\bf d\ell} = \frac{mJ}{q^2}\int_{-a}^a\frac{dz}{\rho},
\end{align}
with $\rho$ given in Eq.\,(\ref{eq:rho_JJinsulator}). Solving Eq.\,(\ref{eq:JJ_phase_int}) for $\alpha$ we get
\begin{equation}
    \alpha^2 = \frac{\sqrt{\rho_1\rho_2}}{2a}\sin\varphi,
\end{equation}
or
\begin{equation}\label{eq:JJ_current2}
    J = \frac{\hbar |q|\sqrt{\rho_1\rho_2}}{2ma}\sin\varphi,
\end{equation}
which is exactly the form given by Eq.~(\ref{eq:JJ_current}), with $J_c= \frac{\hbar |q|\sqrt{\rho_1\rho_2}}{2ma}$. Note that, the Josephson current-phase relation can also be obtained from the order parameter equation. For completeness, in Appendix \ref{append:JJ_EoM} we also provide one such derivation, as well as the discussion on how results obtained from the two approaches are equivalent in the limit of thin insulating region, which is the limit we are interested in.
\begin{figure}[H]
    \centering
    \includegraphics[scale=0.27]{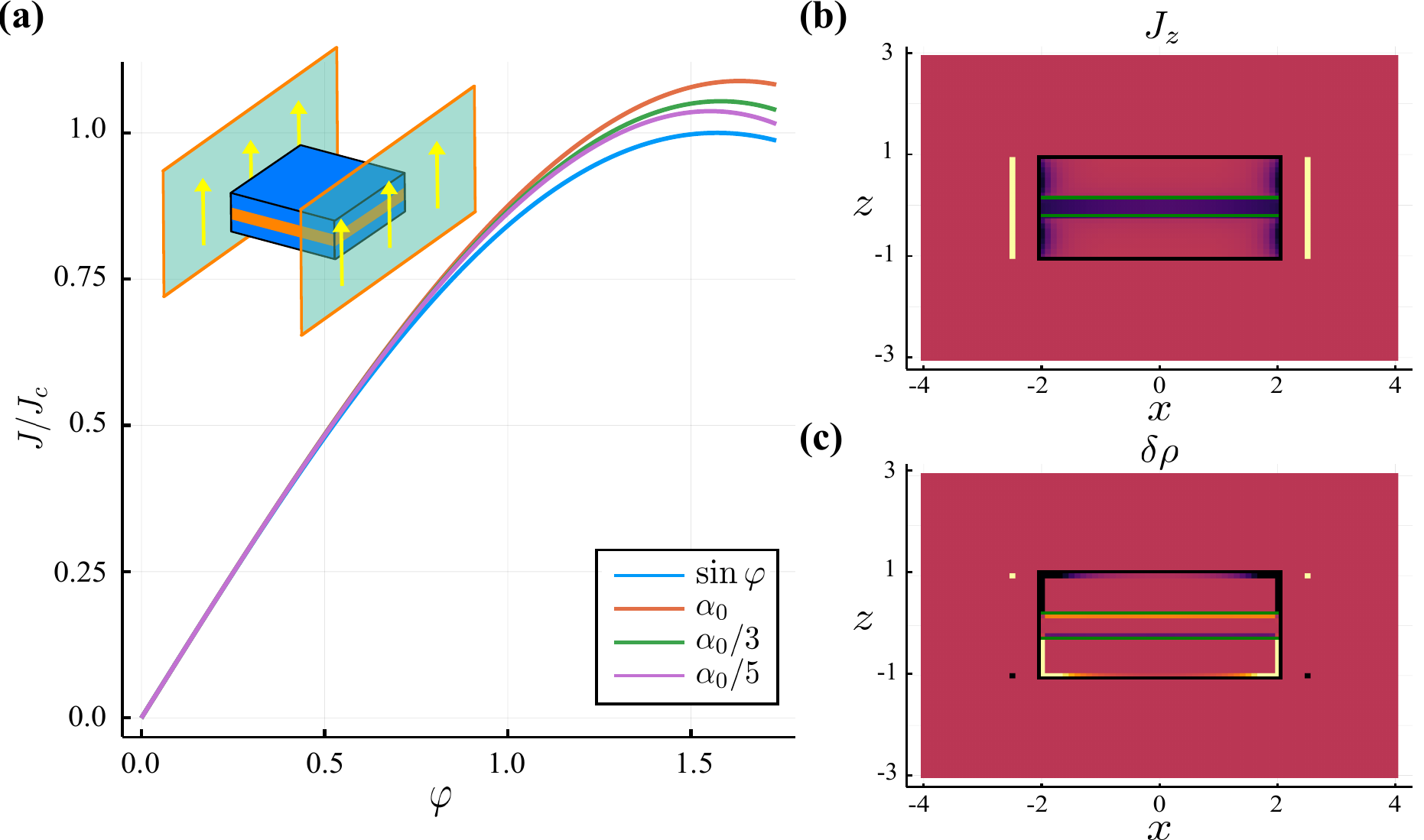}
    \caption{(a) The dependence of current in the insulating region on the phase $\varphi$ is shown. The blue curve is the ideal sinusoidal dependence, and the other colors correspond to different ramping rates of external currents on the surrounding conducting sheets. The inset shows the geometrical set up of the problem. For the two materials that make up the junction, we choose $\lambda_2= 10\lambda_1$. (b) The distribution of current in the junction. (c) The distribution of charge in the junction. All lengths are in units of $\lambda_1$.}
    \label{fig:JJmodel}
\end{figure}

Next, we analyze the ab-initio modeling of the Josephson dynamics through the numerical solution of Eqs.\,(\ref{Eq:DiscreteAmpereLaw_full}) and (\ref{Eq:DiscreteChargeEoM}) in slow dynamics regime. We model the Josephson junction by a sandwich made of two slices of superconductor with a penetration depth $\lambda_1$ and a piece of another superconductor $\lambda_2$ in between, where $\lambda_2\ll \lambda_1$. The $\lambda_2$-superconductor plays the role of the insulating region in our model. The sandwich needs to be thin so that its thickness is much smaller than its lateral dimensions. We then excite the junction by placing it in the gap between two sheets of conductor that are placed parallel to the longitudinal axis of the Josephson sandwich, as shown in the inset of Fig.\,(\ref{fig:JJmodel}a). An external source of magnetic field is created by ramping the uniform current on the sheets. To mimic an adiabatic process, the rate at which the currents on the sheets are ramped up is chosen to be slow compared to the plasma frequency $\omega_j$ of the junction.  

The dependence of current $J$ on the phase $\varphi$ across the insulating region is reported for different rates at which the currents on the conducting sheets increase in Fig.\,(\ref{fig:JJmodel}a), where we define $\alpha_0 = 0.1\omega_j$ to be an arbitrarily slow rate of ramp. The different $J(\varphi)$ plots are compared with the ideal relation $\sin\varphi$. We see that slower rate leads to better agreement with the sinusoidal dependence. This justifies the assumption we made earlier about slow dynamics when deriving the Josephson current-phase relation.

\begin{figure}[H]
    %\centering
    \includegraphics[scale=0.25]{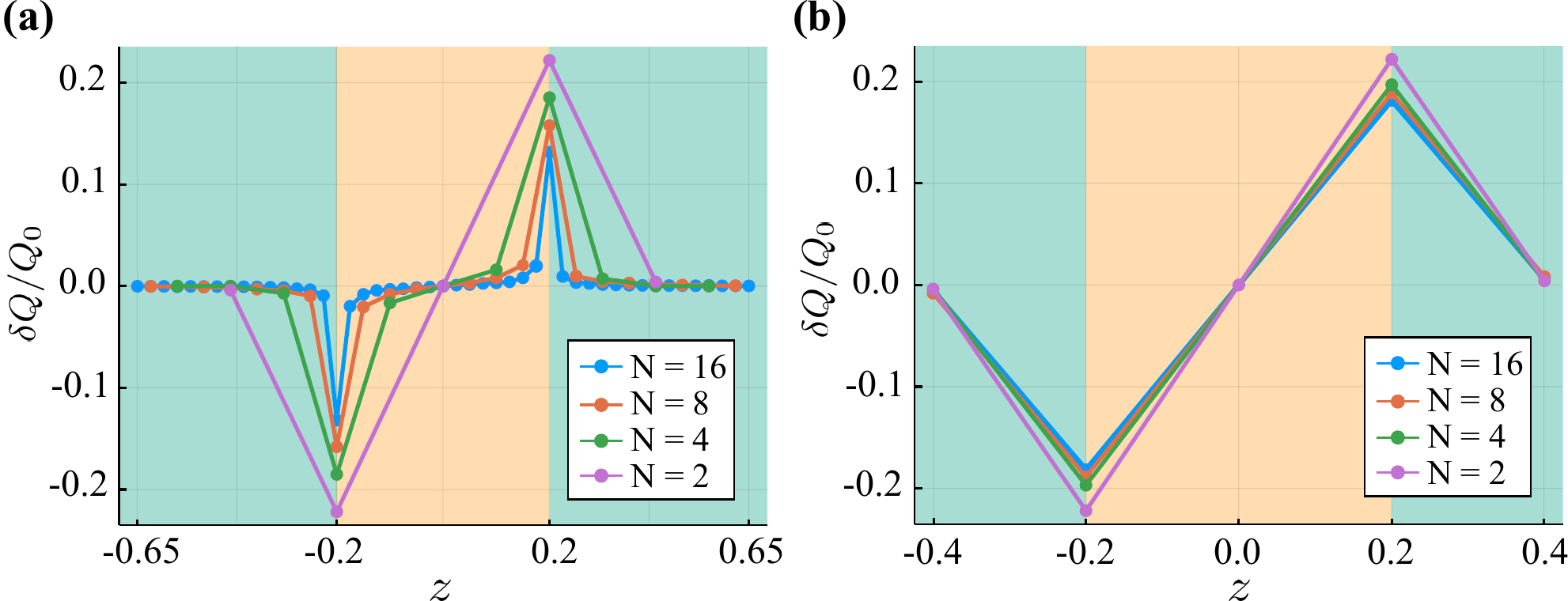}
    \caption{(a) The distribution of charge around the insulating region of the Josephson junction. The yellow shade is the insulator, and the green shades are the superconducting islands. Each colored plot corresponds to a different number of edges $N$ along $z$ used to model the width of the insulating region. (b) The results obtained from different levels of graininess are directly compared by grouping the charge on vertices into `bins' of charges so that the graininess is similar to one would get in the $N=2$ case.}
    \label{fig:coarse_grain}
\end{figure}

We also study the distribution of current and charge in the junction during its interactions with external fields. As seen in Fig.\,(\ref{fig:JJmodel}b), the current flows rather uniformly inside the insulator, 
a result that is largly due to our set up of the problem, where thickness of the insulator is thin enough, and the sheets of current are placed symmetrically on both side of the junction. In addition to the current in the insulator, there are also surface currents that build up on the sides to screen out external magnetic field from entering the superconductor. The charge distribution that stabilizes near the junction, shown in Fig.\,(\ref{fig:JJmodel}c), is the result of a uniform current that flows across the insulator. Charge is therefore drawn from one superconducting island and builds up at the other island, resulting in identical charge distributions of opposite sign at the two interfaces.

The imbalance of charge distribution, manifested through the fluctuation of $\delta\rho$ on top of the uniform background $\rho_0$, happens mostly at material interfaces and vanishes quickly away from boundaries. The impact of such fluctuations on the overall dynamics of the system can be effectively captured by our coarse-grained model.
We computed the distribution of charge around the insulating region with different levels of discretization, indicated by the number of edges $N$ used to model the insulator part (see Fig.\,(\ref{fig:coarse_grain}a)). To directly compare the results obtained from different levels of graininess, the computed charges in each case are lumped into ``bins" of charges so that the resulting distributions have the same spacing as the $N=2$ case (Fig.\,(\ref{fig:coarse_grain}b)). 
The resulting values at the insulator boundaries can then be thought of as the accumulated charges due to the Josephson current. We observe a good agreement between results obtained from grainy ($N=2, 4$) and more fine-grained calculations ($N=8,16$).

\subsubsection{Josephson junction under finite-frequency driving}

The hybridized equations also allow us to study plasma oscillations of a junction. Consider a situation where the junction is driven by an AC source. Performing the integral $\int_1^2(\ref{eq:Aprime_waveeq_rho}) \cdot {\bf d}{\bf \ell}$ across the insulating region of the junction, we obtain
\begin{align}
    \frac{\Phi_0}{2\pi c^2}\frac{\partial^2\varphi}{\partial t^2} =  -\mu_0 J - \frac{q}{2mc^2}\frac{\partial}{\partial t}&\bigg(|A'|^2_2 - |A'|^2_1 \nonumber\\
    &- \frac{\hbar^2}{q^2}\frac{1}{\sqrt{\rho}}\frac{\partial^2\sqrt{\rho}}{\partial z^2}\bigg\rvert_1^2 \bigg).
\end{align}
Combined with Eq.\,(\ref{eq:supercurrent_def}) we obtain
\begin{align}\label{Eq:eom_varphi1}
     \frac{\Phi_0}{2\pi c^2}\frac{\partial^2\varphi}{\partial t^2} =  -\mu_0 J - 
    \frac{\hbar^2}{2mc^2q}\frac{\partial}{\partial t}\bigg(\alpha^4\frac{1}{\rho^2} - \frac{1}{\sqrt{\rho}}\frac{\partial^2\sqrt{\rho}}{\partial z^2}\bigg)\bigg\rvert_1^2.
\end{align}

\begin{figure}[H]
    \centering
    \includegraphics[scale=0.28]{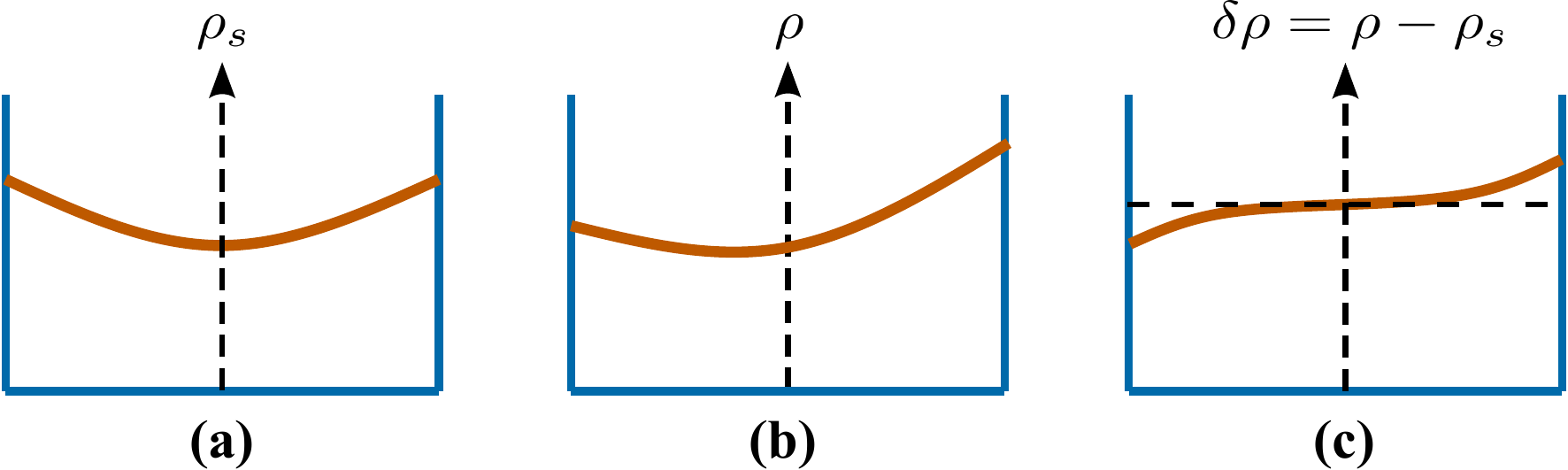}
    \caption{A schematic of condensate distribution in a Josephson junction in (a) steady state, (b) AC driven, and (c) difference between (b) and (a)}
    \label{fig:JJcharge_driven}
\end{figure}
To proceed, we consider the distribution of the condensate $\rho$ in the insulator. At steady state, $\rho = \rho_{s}$, which is given by Eq.\,(\ref{eq:rho_JJinsulator}) and is symmetric about $z=0$ as seen in Fig.\,(\ref{fig:JJcharge_driven}a). Now consider the situation where the junction is slowly driven by an AC source, and at time $t$ the condensate becomes imbalanced as shown in Fig.\,(\ref{fig:JJcharge_driven}b). We consider a weak driving case so that the fluctuation of charge $\delta\rho$ introduced is small ($\delta\rho\ll\rho_s$). Based on the symmetries of $\rho_s$ and $\delta\rho$, we then have the following properties
\begin{align}\label{Eq:JJrho_properties}
    \rho_{s}\rvert_1 = \rho_{s}\rvert_2 &, \hspace{0.1in} \delta\rho\rvert_1 = -\delta\rho\rvert_2 \nonumber\\
    \frac{d\rho_s}{dz}\Big\rvert_1 = -\frac{d\rho_s}{dz}\Big\rvert_2 &, \hspace{0.1in} \frac{d\delta\rho}{dz}\Big\rvert_1 = \frac{d\delta\rho}{dz}\Big\rvert_2 \\
    \frac{d^2\rho_s}{dz^2}\Big\rvert_1 = \frac{d^2\rho_s}{dz^2}\Big\rvert_2 &, \hspace{0.1in} \frac{d^2\delta\rho}{dz^2}\Big\rvert_1 = -\frac{d^2\delta\rho}{dz^2}\Big\rvert_2 \nonumber.
\end{align}
We now expand the terms in Eq.\,(\ref{Eq:eom_varphi1}) to first order in $\delta\rho$
\begin{align}
    \frac{1}{\sqrt{\rho}}\frac{\partial^2\sqrt{\rho}}{\partial z^2} &= \frac{1}{2\rho}\frac{\partial^2\rho}{\partial z^2} - \frac{1}{4\rho^2} \Big(\frac{\partial\rho}{\partial z}\Big)^2 \\
    &\approx \frac{1}{2\rho_s}\bigg[\frac{d^2\rho_s}{dz^2} + \frac{\partial^2\delta\rho}{\partial z^2} - \bigg(\frac{d^2\rho_s}{dz^2} \bigg)\frac{\delta\rho}{\rho_0}\bigg] \nonumber\\
    & - \frac{1}{4\rho_0^2}\bigg[\bigg(\frac{d\rho_s}{dz} \bigg)^2\bigg(1- \frac{2\delta\rho}{\rho_s}\bigg) + 2\frac{d\rho_s}{dz}\frac{\partial\delta\rho}{\partial z} \bigg] \nonumber.
\end{align}
We can take advantage of the properties shown in Eqs.\,(\ref{Eq:JJrho_properties}) to obtain
\begin{align}
    \frac{1}{\sqrt{\rho}}\frac{\partial^2\sqrt{\rho}}{\partial z^2}\bigg\rvert_1^2 = \frac{1}{2\rho_s}\bigg[\frac{d^2\rho_s}{dz^2} -& \bigg(\frac{d^2\rho_s}{dz^2} \bigg)\frac{\delta\rho}{\rho_s} + \frac{\delta\rho}{\rho_s^2}\bigg(\frac{d\rho_s}{dz}\bigg)^2  \nonumber\\
    & - \frac{1}{\rho_s}\frac{d\rho_s}{dz}\frac{\partial\delta\rho}{\partial z} \bigg]\bigg\rvert_1^2.
\end{align}
To proceed, we now consider the driving frequency to be in the RF regime. Since the wavelength is of order $mm$, many orders of magnitude longer than the typical size of the insulator (a few $nm$s), the electromagnetic field felt by the junction is much smaller than the wavelength. Based on this observation, we consider an ansatz for $\delta\rho$ in the insulator, $\delta\rho = \beta(z,t)z$, with the lowest order being $\beta(z,t) = \beta(t)$. This corresponds to the assumption that there is a weak and slow AC drive that slightly tilts the distribution of the condensate $\rho$. A natural question that arises is whether instead of a linear tilt, $\rho$ should change sharply at the insulator interfaces. This is indeed true, but the sharpness of $\rho$ is exhibited in the DC distribution $\rho_s$. The fluctuation $\delta\rho$ due to AC driving, however, need not be sharp. The expansion of $\rho$ into a static part and a fluctuating smaller part $\delta \rho$ allows us to decouple these two effects. We can then rewrite Eq.\,(\ref{Eq:eom_varphi1}) as follows
\begin{align}
    \frac{\partial^2\varphi}{\partial t^2} =  -\bigg(\frac{2\pi \mu_0c^2}{\Phi_0}\bigg) J +& \frac{4\pi m}{\hbar q^2\rho_s^3}\frac{d(J^2\delta q_2)}{dt} \nonumber\\
    + \frac{\hbar^2}{2mc^2q\rho_s}\frac{d}{dt}\bigg\{\bigg[\frac{1}{\rho_s}&\bigg(\frac{d\rho_s}{dz}\bigg)^2  -\bigg(\frac{d^2\rho_s}{dz^2} \bigg) \bigg]\frac{\delta\rho_2}{\rho_s} \nonumber\\
    &-\frac{1}{\rho_s}\frac{d\rho_s}{dz}\frac{\partial\delta\rho}{\partial z}\bigg\rvert_2\bigg\}, 
\end{align}
where $\delta\rho_2=\delta\rho(a)$. Plugging in the ansatz for $\delta\rho$, we get
\begin{align}\label{Eq:eom_varphi_beta}
    \frac{\partial^2\varphi}{\partial t^2} =  -\bigg(\frac{2\pi \mu_0c^2}{\Phi_0}\bigg) J +& \frac{4\pi ma}{\hbar q^2\rho_2^3}\frac{d(J^2\beta)}{dt} \nonumber\\
    + \frac{\hbar^2}{2mc^2q\rho_2}\bigg\{\bigg[\frac{1}{\rho_2}&\bigg(\frac{d\rho_s}{dz}\bigg)^2  -\bigg(\frac{d^2\rho_s}{dz^2} \bigg) \bigg]\frac{a}{\rho_2} \nonumber\\
    &-\frac{1}{\rho_2}\frac{d\rho_s}{dz}\bigg\}\bigg\rvert_2\frac{d\beta}{dt}. 
\end{align}
We can relate the slope $\beta(t)$ to the number of additional Cooper pairs $n$ that accumulate at one side of the junction (with the other side losing equal number $n$ of pairs)
\begin{equation}
    n = \frac{1}{8}\beta a^2 A_{J},
\end{equation}
where $A_{J}$ is the cross section area of the junction. The charge conservation law reads
\begin{equation}
    q\frac{dn}{dt} = JA_J,
\end{equation}
where $J$ is the current density at the symmetry point $z=0$, which in lowest order can be taken to be equal to the DC Josephson current given in Eq.\,(\ref{eq:JJ_current2}). 
It is also convenient to write the equation of motion in terms of $n$ instead of $\delta\rho$ and  Eq.\,(\ref{Eq:eom_varphi_beta}) becomes
\begin{align}\label{eq:varphi2ndorder}
    \frac{\partial^2\varphi}{\partial t^2} &=  -\bigg(\frac{2\pi \mu_0c^2}{\Phi_0}\bigg) J_c\sin\varphi + \frac{32\pi mJ_c^2}{\hbar q^2 a A_J\rho_2^3}\bigg(\sin^2\varphi\frac{dn}{dt} \nonumber\\
    &+ 2\sin\varphi\cos\varphi\frac{d\varphi}{dt}n \bigg) + \frac{4\hbar^2}{A_Jmc^2a^2q\rho_2}\bigg\{\bigg[\frac{1}{\rho_2}\bigg(\frac{d\rho_s}{dz}\bigg)^2 \nonumber\\
    &-\bigg(\frac{d^2\rho_s}{dz^2} \bigg) \bigg]\frac{a}{\rho_2} -\frac{1}{\rho_2}\frac{d\rho_s}{dz}\bigg\}\bigg\rvert_2\frac{dn}{dt}.
\end{align}
It is useful to introduce the definition of junction energy, which given by 
\begin{equation}
    \mathcal{E}(t) = \int_0^t\!Iv\,dt = E_j\big[\cos(\varphi(0))-\cos\varphi\big],
\end{equation}
where $E_j=\hbar J_c A_J/q$ and $v=\int_1^2{\bf d\ell}\cdot{\bf E}$. Consider the following term in Eq.\,(\ref{eq:varphi2ndorder})
\begin{align}\label{eq:intbypart}
    \frac{\partial\varphi}{\partial t}n &= \bigg(\frac{q}{\hbar}\int_1^2\!\!{\bf d\ell}\cdot{\bf E}\bigg)\bigg(\frac{1}{q}\!\int_0^t\!\!Idt\bigg) \\
    &= \frac{1}{\hbar}\bigg(\int_0^t\!\!Ivdt  - I\!\!\int_0^t\!\!vdt\bigg) \nonumber\\
    &= \frac{1}{\hbar}\bigg(\mathcal{E}(t) - I\frac{\hbar}{q}(\varphi(t) - \varphi(0)) \bigg) \nonumber\\
    &= \frac{J_cA_J}{q}\big[\cos(\varphi(0))-\cos\varphi - (\varphi(t)-\varphi(0))\sin\varphi\big]. \nonumber
    \end{align}
Now combining all the pieces together, the equation of motion for $\varphi$ is given by
\begin{widetext}
\begin{align}\label{eq:JJoscillation}
    \frac{\partial^2\varphi}{\partial t^2} =  -\Bigg(& \frac{2\pi \mu_0 q c^2}{\hbar} - \frac{4\hbar^2}{mc^2a^2q^2\rho_2}\bigg\{\bigg[\frac{1}{\rho_2}\bigg(\frac{d\rho_s}{dz}\bigg)^2 -\bigg(\frac{d^2\rho_s}{dz^2} \bigg) \bigg]\frac{a}{\rho_2} -\frac{1}{\rho_2}\frac{d\rho_s}{dz}\bigg\}\bigg\rvert_2 
    \Bigg) J_c\sin\varphi \\
    & + \frac{32\pi mJ_c^3}{\hbar a q^3  \rho_2^3}\sin\varphi\Big[\sin^2\varphi + 2\cos\varphi\big(\cos(\varphi(0))-\cos\varphi - (\varphi(t)-\varphi(0))\sin\varphi\big) \Big]. \nonumber
\end{align}
\end{widetext}
In Eq.\,(\ref{eq:JJoscillation}), the terms on the first line represent the standard Josephson plasma oscillation. The first term in the big bracket on RHS gives the standard Josephson plasma frequency, while the second term in the bracket provides the correction to that frequency. The term on the second line is a correction beyond the standard Josephson plasma oscillation featuring higher harmonics. Interestingly, these corrections contain a term that is not periodic in the phase $\varphi$.  
\begin{figure}[H]
    %\centering
    \includegraphics[scale=0.25]{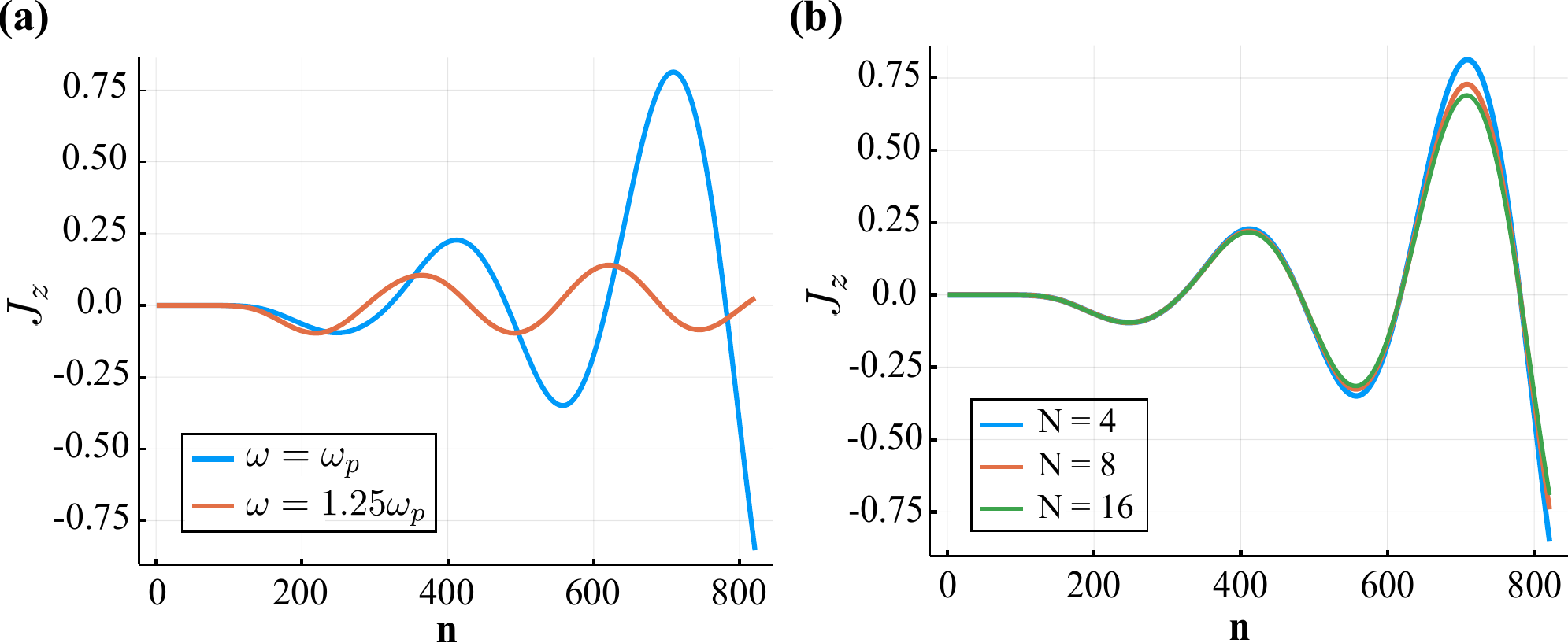}
    \caption{(a) The current density over time at the center of the insulator. The blue curve corresponds to the resonant case, when the junction is driven at its plasma frequency $\omega_p$, while the blue curve corresponds to when driving is off resonance. (b) Current density when the junction is driven at resonance. Each color corresponds to a different number of edges $N$ along $z$ used to model the width of the insulating region.}
    \label{fig:coarse_grain_AC}
\end{figure}
Using the same toy model for the JJ as the one used in calculations above, we can simulate the resulting dynamics when the JJ is driven by an oscillating field. We perform a frequency sweep around the range that is expected to find the junction plasma frequency. With a sampling rate of $\Delta\omega/\omega\approx 5\%$ we found that three choices of $N \in \{4,8,16\}$ yields the same plasma frequency $\omega_p$. The responses of the junction when driven at and off resonance are compared and shown in Fig.\,(\ref{fig:coarse_grain_AC}a), where the current density over time at the center of the insulator is plotted. The at-resonance current density computed using different values of $N$ is shown in Fig.\,(\ref{fig:coarse_grain_AC}b), where we see good convergence of the three curves.

\section{Discussion and Conclusions}

In this paper, we introduce a computational approach for solving the equations that govern the dynamics of the order parameter of three-dimensional superconducting materials interacting with electromagnetic fields. To achieve this, we solve the non-linear Schr\"odinger equation, which describes the dynamics of the order parameter and the Maxwell's equations, which describe the dynamics of the electromagnetic fields. While these equations have been previously written, the new contributions of this paper are as follows:

(1) We present a systematic method for numerically solving the equations for the coarse-grained electromagnetic and charge degrees of freedom. In this approach, we introduce a method for compressing resources while maintaining accuracy for the coarse-grained fields. This is all that is needed because, in any physical measurement, the acquired results are always coarse-grained. The meshing need not be finer than the resolution of these channels or the size of the JJs. This insight is helpful to the designing of a cleverly employed adaptive meshing strategy to reduce computational demand. As such, this approach provides the first step towards a general systematic method for optimizing the balancing between accuracy and resources of the numerical solution of non-linear spatio-temporal partial differential equations. We hope to demonstrate this further in future works.

(2) We provide equations that operate on gauge-invariant hybridized variables of electromagnetic and charge degrees of freedom. In general, the expression of the Maxwell's equations in terms of the electromagnetic fields $\bm{E}$ and $\bm{B}$ render them manifestly gauge-invariant. However, the quantization of light-matter interactions typically relies on the minimal coupling form through ($\bm{A}$, $V$)~\cite{Gell-Mann_qed}, in which the gauge must be fixed. In this paper, we take a different approach and write the known minimal-coupling form of the order parameter equation of a BCS superconductor in terms of the fields ($\mb{A}'$, $\rho$), which are manifestly gauge-invariant. This allows for the second quantization of these equations along a similar vein as the standard approach ~\cite{blaisrmp, 1+1D_Moein_2016} and in
a gauge-invariant manner. We plan to address second quantization of DEC-QED in future work.

(3) We present a set of discretized equations based on DEC. These equations provide the geometric scaffolding necessary for the numerical solution of the resulting equations, ensuring stability and accuracy in the long-term and over large spatial regions. Without this scaffolding, the solution of non-linear equations can often display instabilities in the long-time limit. By contrast, the methods introduced in this paper may alleviate such issues.

Our analysis is missing a coarse-graining procedure in the time-domain, which is necessary for maintaining relativistic invariance. In addition, further work is needed to improve the coarse-graining of the non-linear terms and to understand their impact on the numerical accuracy of the solutions. These improvements will likely require the development of new mathematical tools and techniques. Additionally, detailed analysis of the numerical stability of the equations presented is needed and will be left for future work.

Moreover, previous research in the context of 1+1D cQED theory for transmission lines has shown that open boundary conditions can be applied to second quantized macroscopic fields in order to accurately capture radiative losses to and thermalization with the electromagnetic vacuum surrounding the finite volume~\cite{kanupaper, cuttoff_free_cqed_2017}. In order to make DEC-QED a useful tool for simulating the quantum dynamics of 3D superconducting devices, it is necessary to extend this formulation to DEC-QED.

\section{Acknowledgements}
We gratefully acknowledge discussions with Anil N. Hirani, Saeed A. Khan, Richard Li, Thomas Maldonado, Alejandro Rodriguez, Kanu Sinha, and Hoang Le. We acknowledge support from the US Department of Energy, Office of Basic Energy Sciences, Division of Materials Sciences and Engineering, under Award No. DESC0016011. The simulations presented in this article were performed on computational resources managed and supported by Princeton Research Computing, a consortium of groups including the Princeton Institute for Computational Science and Engineering (PICSciE) and the Office of Information Technology's High Performance Computing Center and Visualization Laboratory at Princeton University.\\

%\newpage
\appendix
\section{Maxwell's operators in terms of discrete flux coordinates}\label{append:discreteMaxwell}
We derive a discrete form of the divergence and curl-curl operators that are suitable for our use of flux coordinates. For a vector field ${\bf F}$, an integral of ${\bf \nabla}.{\bf F}$ over the volume of a cell gives
\begin{align}\label{eq:discretediv}
    \int_{v^{\dagger}}(\nabla \cdot {\bf F})d{\bf r}^3 &=  \int_{\partial (v^{\dagger})}\!{\bf F} \cdot {\bf da}\nonumber\\
    &=\sum_{e^{\dagger}\in\partial (v^{\dagger})}\int_{e^{\dagger}}\!{\bf F} \cdot {\bf da}\nonumber\\
    &= \sum_{e^{\dagger}\in\partial (v^{\dagger})}\langle F_{\perp e^{\dagger}}\rangle_{e^{\dagger}} \cdot  \Delta A(e^{\dagger})\nonumber\\
    &\approx \sum_{e|v\in e}\langle F_{\perp e^{\dagger}}\rangle_{e}  \cdot  \Delta A(e^{\dagger}) \nonumber\\
    &= \sum_{e|v\in e} \frac{\Delta A(e^{\dagger})}{\Delta\ell(e)}\int_e {\bf F} \cdot {\bf d\ell}, 
\end{align}
where $\Delta A(e^{\dagger})$ is the area of the face $e^{\dagger}$, $\Delta\ell(e)$ is the length of the edge $e$,  and $F_{\perp e^{\dagger}}$ is the component of ${\bf F}$ normal to the face $e^{\dagger}$. Here $\langle F_{\perp e^{\dagger}}\rangle_{e^\dagger} = \frac{1}{\Delta A(e^\dagger)}\int_{e^\dagger}{\bf F}\cdot{\bf da}$ is the average of $F_{\perp e^{\dagger}}$ over the area of $e^{\dagger}$. Similarly $\langle F_{\perp e^\dagger}\rangle_{e}$ is the average of the same field component over the edge $e$ dual to $e^{\dagger}$. In Eq.\,(\ref{eq:discretediv}) above, we have made an approximation by assuming that $\langle F_{\perp e^{\dagger}}\rangle_{e^\dagger} = \langle F_{\perp e^\dagger}\rangle_{e}$. This procedure is formally called a Hodge star operation, where a transformation of a field living on the dual mesh to a field that lives on the primal mesh, and vice versa \cite{diff_form_book_Flanders}. The approximation comes in through the discrete representation of the Hodge star. The error of this operation converges in first order in the characteristic length of the grid for scalar fields and in second order for vector fields \cite{hirani_hodge_convergence}.

Next, for an edge $e\in M$, the discrete curl-curl operator is given by
\begin{align}
    \int_{e^{\dagger}}\! (\nabla\!\times\!\nabla\!\times {\bf F}) \cdot {\bf da} &= \int_{\partial(e^{\dagger})}(\nabla\!\times {\bf F}) \cdot {\bf d\ell} \\
    &= \sum_{e_0\in\partial(e^{\dagger})}\int_{e_0}(\nabla\!\times {\bf F})  \cdot {\bf d\ell} \nonumber\\
    &= \sum_{e_0\in\partial(e^{\dagger})} \langle \nabla\!\times {\bf F} \rangle_{e_0} \cdot \Delta\ell(e_0)\nonumber\\
    &\approx \sum_{e_0\in\partial(e^{\dagger})} \frac{\Delta\ell(e_0)}{\Delta A(e_0^{\dagger})} \int_{e_0^{\dagger}}\!(\nabla\!\times {\bf F}) \cdot {\bf da} \nonumber\\
    &= \sum_{e_0\in\partial(e^{\dagger})} \frac{\Delta\ell(e_0)}{\Delta A(e_0^{\dagger})} \sum_{e_1\in\partial(e_0^\dagger)}\int_{e_1}\!{\bf F} \cdot {\bf d\ell}. \nonumber
\end{align}

We also derive here the full form of the nonlinear terms whose graphical representations are used in Eq.\,(\ref{Eq:DiscreteAmpereLaw_full}). 
\begin{widetext}

Consider an edge $e$ bounded by the vertices $[v_A, v_B]$ and oriented from $v_A$ to $v_B$, the action of the operator $\gradAsq$ on $|{\bf A'}|^2$ evaluated at $e$ is given by
\begin{align}
    \int_{e^\dagger} \nabla\big|{\bf A'}\big|^2\!\cdot{\bf da} &= \frac{\Delta A(e^\dagger)}{\Delta\ell(e)}\Big[{\bf A}'|^2(v_B) - |{\bf A}'|^2(v_A)\Big] \\
    &= \frac{\Delta A(e^\dagger)}{\Delta\ell(e)}\bigg[\sum_{e_0\supset v_B} \frac{\Delta V(v_B^{\dagger})\cap V_{s}(e_0)}{\Delta V(v_B^{\dagger})^2}\frac{\Delta A(e_0^\dagger)} {\Delta\ell(e_0)}\Phi(e_0)^2 
    - \sum_{e_0\supset v_A} \frac{\Delta V(v_A^{\dagger})\cap V_{s}(e_0)}{\Delta V(v_A^{\dagger})^2}\frac{\Delta A(e_0^\dagger)} {\Delta\ell(e_0)}\Phi(e_0)^2 \bigg], \nonumber
\end{align}
where $V_s(e_0)$ is the support volume of the edge $e_0$ \cite{DEC_HiraniThesis}. Simillarly, the full form of the quantum pressure term evaluated at $e$ is given by
\begin{align}
    \int_{e^\dagger}\nabla\bigg[\frac{\nabla^2(\sqrt{\rho})}{\sqrt{\rho}}\bigg]\!\cdot{\bf da} = \frac{\Delta A(e^\dagger)}{\Delta\ell(e)} & \bigg[\frac{\nabla^2(\sqrt{\rho})}{\sqrt{\rho}}\bigg\rvert_{v_B} -\frac{\nabla^2(\sqrt{\rho})}{\sqrt{\rho}}\bigg\rvert_{v_A}\bigg] \nonumber\\
    = \frac{\Delta A(e^\dagger)}{\Delta\ell(e)} & \Bigg\{\frac{1}{\Delta V(v_B^\dagger)\rho(v_B)}\sum_{e_0^\dagger\in\partial v_B^\dagger}\frac{\Delta A(e_0^\dagger)}{\Delta\ell(e_0)}\sum_{v_1\in\partial e_0}\Big[\sqrt{\rho}(v_1) -\sqrt{\rho}(v_B)\Big] \nonumber\\
    -& \frac{1}{\Delta V(v_A^\dagger)\rho(v_A)}\sum_{e_0^\dagger\in\partial v_A^\dagger}\frac{\Delta A(e_0^\dagger)}{\Delta\ell(e_0)}\sum_{v_1\in\partial e_0}\Big[\sqrt{\rho}(v_1) -\sqrt{\rho}(v_A)\Big]\Bigg\}.
\end{align}

\end{widetext}

\section{The Reduction to the 1+1D Theory}\label{append:1Dderivations}
\begin{figure}[h!]
    \centering
    \includegraphics[scale=0.45]{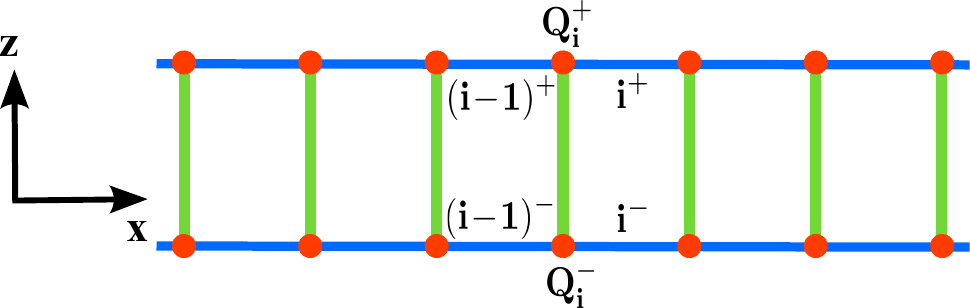}
    \caption{Discretization in a 1D waveguide}
    \label{fig:1d_grid}
\end{figure}
In this section we show that the 1+1D theory of cQED~\cite{blaisrmp,1+1D_Moein_2016} can be recovered by considering a limiting case of the full 3+1D flux-based electromagnetic theory. Specifically, we seek to reproduce the equation of motion for the flux in a 1D waveguide, which in standard cQED is usually derived using the lumped-element circuit model. We will pay attention to the derivation of the effective parameters of the 1+1D reduced equations from the 3+1D formulation. 

Consider a one-dimensional waveguide along $x$ made up of two parallel superconducting plates with a spacing $d$ between them in $z$ direction. We assume translational invariance in material properties along $y$
and assume there is no variation of the fields in that direction. The discretization of space inside the waveguide (plus the boundaries) is shown in Fig.\,(\ref{fig:1d_grid}).
The region between the plates can be assumed to be vacuum (or air) with permitivity $\epsilon$ and Permeability $\mu$. Within this region there is no superelectrons and hence the phase $\theta$ takes up an arbitrary constant value. In our one-dimensional model, this region is discretized by one layer of equally spaced vertical edges that spans the entire $x$-axis. The two superconducting plates that form the top and bottom boundaries of the waveguide are subdivided into finite horizontal edges. Charges $Q_i^{\pm}$ are placed at the vertices on these plates, while fluxes $\Phi_{x,z}$ are defined on the edges. 
Eq.\,(\ref{Eq:DiscreteAmpereLaw_full}), when applied to the $i^{th}$ vertical edge, then reduces to
\begin{align}
    \mu\varepsilon\partial^2_t{\Phi}_z^i\!-\!\partial_x^2{\Phi}_z^i\!+\!\frac{ ({\Phi}_x^{i^+}\!-\!{\Phi}_x^{i^-}) - ({\Phi}_x^{(i-1)^+}\!\!-\!{\Phi}_x^{(i-1)^-})}{\Delta \ell_x^2}& \nonumber\\ =-\mu\varepsilon\dot{\psi}_z^i,&
\end{align}
where $\Delta\ell_x$ is the spacing between two consecutive vertical edges. We can further simplify the equation by assuming an initial condition such that there is charge balance between the upper and lower superconducting surfaces. Then charge at nodes $i^+$ and $i^-$ satisfies $Q_i^- =-Q_i^+$, meaning that the current at the lower plane flows in opposite direction as the current in the upper plane. Therefore $\phi_x^{i^-} = -\phi_x^{i^+}$, and Ampere's law becomes
\begin{equation}\label{eq:Ampere_1Dres}
    \mu\varepsilon\partial^2_t{\Phi}_z^i - \partial_x^2{\Phi}_z^i - \frac{ 2({\Phi}_x^{i^+} - {\Phi}_x^{(i-1)^+})}{\Delta \ell_x^2} = - \mu\varepsilon\dot{\psi}_z^i.
\end{equation}
Gauss's law at the node $Q_i^+$ is given by
\begin{align}\label{eq:Gauss_1Dres}
    \frac{Q_i^+}{\epsilon} &= -\frac{\Delta\ell_z}{\Delta\ell_x}(\psi_x^{i^+}-\psi_x^{(i-1)^+}) + \frac{\Delta\ell_x}{\Delta\ell_z}\psi_z^i, \nonumber\\
    &\approx \frac{\Delta\ell_x}{\Delta\ell_z}\psi_z^i,
\end{align}
where $Q_i^+$ is the charge per unit length along $y$, and we have assumed that in the continum limit, $\Delta\ell_x\rightarrow 0$, the charge distribution is smooth along $x$. Hence $|\psi_x^{i^+}-\psi_x^{(i-1)^+}|\ll |\psi_z^i|$. The supercurrent along $x$ at the surface of the superconductor is given by
\begin{equation}\label{eq:supercurrent}
    J_x = -\frac{{\Phi}_x}{\mu\lambda_L^2\Delta\ell_x}.
\end{equation}
Differentiating in time Eq.\,(\ref{eq:Gauss_1Dres}) and using Eq.\,(\ref{eq:supercurrent}) to represent $\dot{Q}_i^+=(J_x^{(i-1)^+}\!-J_x^{i^+})\Delta\ell_z$, we obtain 
\begin{equation}\label{eq:1d_gauss}
    \frac{{\Phi}_x^{i^+}-{\Phi}_x^{(i-1)^+}}{\Delta\ell_x^2} = \frac{\epsilon\mu\lambda_L^2}{d^2}\dot{\psi_z^i}.
\end{equation}
Combining this with Eq.\,(\ref{eq:Ampere_1Dres}), we get
\begin{equation}\label{eq:wave_eq_phiZ}
    \mu\epsilon\partial_t[\partial_t{\Phi}_z^i + \psi_z^i] - \partial_x^2{\Phi}_z^i - 2\mu\epsilon\frac{\lambda_L^2}{d^2}\partial_t\psi_z^i = 0.
\end{equation}
On the other hand, the wave equation for the electric field in vacuum with no charge is given by
\begin{equation}\label{eq:wave_eq_Ez}
    \mu\epsilon\partial^2_tE_z - \partial_x^2E_z = 0.
\end{equation}
Performing (\ref{eq:wave_eq_phiZ}) \!$-$\! $\int\!dt$(\ref{eq:wave_eq_Ez}), we get
\begin{equation}\label{eq:wave_eq_psiz}
    \partial_x^2\Psi_z^i - 2\mu\epsilon\frac{\lambda_L^2}{d^2}\partial_t^2\Psi_z^i = 0,
\end{equation}
where $\Psi_z^i = \int\!\psi_z^idt$. For the system we are considering $c= L_y\frac{\epsilon}{d}$ and $l=\frac{2\mu\lambda_L^2}{L_yd}$, with $L_y$ being the size of the waveguide in $y$ direction. Therefore Eq\,(\ref{eq:wave_eq_psiz}) is equivalent to Eq.\,(\ref{eq:wave_eq_1D_Psiz}). 

We are also interested in the details in how our discrete formulation of Maxwell electrodynamics relates to the transmission line circuit theory in a physical sense. Consider the flux through the $i^{\text{th}}$ cell in the $xz$ plane
\begin{equation}\label{eq:fluxthroughcell}
    \varphi_i = {\Phi}_x^{i^+}\! + {\Phi}_z^i - {\Phi}_z^{i+1} - {\Phi}_x^{i^-} \approx {\Phi}_x^{i^+}- {\Phi}_x^{i^-} = 2{\Phi}_x^{i^+}. 
\end{equation}
On the other hand, from Eq.\,(\ref{eq:1d_gauss}) we get
\begin{equation}
    \frac{{\Phi}_x^{i^+}\!\!-{\Phi}_x^{(i-1)^+}}{\Delta\ell_x} \!= \frac{\epsilon\mu\lambda_L^2}{d^2}\Delta\ell_x\dot{\psi_z^i} = \frac{1}{2}\partial_x^2\Big(\!\int\!\! \psi_z^idt\Big) \!= \frac{1}{2}\partial_x^2\Psi_z^i,
\end{equation}
where in the last expression above we have used Eq.\,(\ref{eq:wave_eq_1D_Psiz}). From finite difference form at the LHS of the equation above going to the continum limit, we get
\begin{equation}
    \partial_x{\Phi}_x = \frac{1}{2}\Delta_x\partial_x^2\Psi_z^i.
\end{equation}
Integrating both side, we get
\begin{equation}
    2{\Phi}_x = \Delta\ell_x\partial_x\Psi_z^i = \Delta_x\Psi_z^i = \Psi_z^i - \Psi_z^{i-1}.
\end{equation}
Combining with the result in Eq.\,(\ref{eq:fluxthroughcell}) the equation above gives us
\begin{equation}
    \varphi_i = \Psi_z^i - \Psi_z^{i-1}.
\end{equation}
Therefore, the flux through a unit cell in the 1D transmission line waveguide is equal to the difference of $\Psi$ defined at the two nodes of a cellular inductor in the lumped-element circuit. 

\section{Dynamics of the Josephson junction}\label{append:JJ_EoM}

In this appendix, we provide a derivation of DC Josephson effect starting from the order parameter equation (Eq.\,(\ref{eq:generalSE})). We also show how the resulting equation is equivalent to what was obtained in Section \ref{sec:JJdynamics}, which starts from the hybridized field equation in Eq.\,(\ref{Eq:DiscreteAmpereLaw_full}). 
We first solve the order parameter equation for the condensate wavefunction in the insulating region $[-a,a]$ in the case of ${\bf A}=0, V=0$. Assuming an ansatz $\Psi(z,t) = \Phi(z)e^{-iEt/\hbar}$ for the wavefunction, from Eq.\,(\ref{eq:generalSE}) we get
\begin{equation}\label{eq:Psi0_insulator}
\Psi(z,t) = C\cosh{(\kappa z)} + D\sinh{(\kappa z)},
\end{equation}
with $\kappa = \sqrt{{2m(U_0-E)}/{\hbar^2}}$. Assuming the time-dependent boundary conditions $\Psi(-a,t) = \sqrt{\rho_1}e^{i\theta_1}$, and $\Psi(a,t) = \sqrt{\rho_2}e^{i\theta_2}$ for $z=-a$ and $z=a$ respectively, we obtain
\begin{equation}\label{eq:CD_coefs}
C = \frac{\sqrt{\rho_1}e^{i\theta_1} + \sqrt{\rho_2}e^{i\theta_2}}{2\cosh{(\kappa a)}}, \text{and } D = \frac{\sqrt{\rho_2}e^{i\theta_2} - \sqrt{\rho_1}e^{i\theta_1}}{2\sinh{(\kappa a)}}.
\end{equation}
The current density is then given by
\begin{equation}
J = \frac{q\hbar\kappa}{m}\im\sbkt{C^*D} = \frac{q\hbar\kappa\sqrt{\rho_1\rho_2}}{m\sinh{(2\kappa a)}}\sin(\theta_2-\theta_1).
\end{equation}
Define
\begin{equation}
    J_c = \frac{|q|\hbar\kappa\sqrt{\rho_1\rho_2}}{m\sinh{(2\kappa a)}}, \varphi = \theta_1 - \theta_2, 
\end{equation}
then we obtain the usual Josephson current equation, $J = J_c\sin\varphi$. If EM field is present (${\bf A} \neq 0, V\neq0$), then the phase $\varphi$ is given by Eq.\,(\ref{eq:JJvarphi_define}). Now consider the limit $\kappa a\ll 1$, when the thickness $a$ of the insulator is very small. Then $\sinh(2\kappa a)\approx 2\kappa a$, and the critical current becomes
\begin{equation}
    J_c = \frac{|q|\hbar\kappa\sqrt{\rho_1\rho_2}}{m\sinh{(2\kappa a)}} \approx \frac{|q|\hbar\sqrt{\rho_1\rho_2}}{2ma},
\end{equation}
which is the same critical current obtained from our ${\bf A}'$-field formulation. Moreover, we can show that the condensate wavefunction in Eq.\,(\ref{eq:Psi0_insulator}) reduces to the same limit as was achieved in Eq.\,(\ref{eq:rho_JJinsulator}). From Eq.\,(\ref{eq:Psi0_insulator}), we have
\begin{align}
    \rho(z) &= |\Psi^2| \\
            &\approx (CC^* + DD^*)\kappa^2x^2 + (CD^* + C^*D)\kappa x + CC^*. \nonumber
\end{align}
Using the results for $C$ and $D$ in Eq.\,(\ref{eq:CD_coefs}), we obtain a form for $\rho(z)$ that is exactly the same as Eq.\,(\ref{eq:rho_JJinsulator}). Therefore, we have shown that the two approaches achieve the same results for the Josephson current-phase relation.

\section{Perturbative Analysis of Non-linear Equations}\label{append:perturbationtheory}
The non-linearities in Eqs.\,(\ref{eq:Aprime_waveeq}) and (\ref{eq:chargeconserve2}), which in most cases can be treated perturbatively, can be addressed systematically by keeping track of the harmonic orders generated. Considering a source term at frequency $\omega$, we can make the following ansatz for $ \vec A'$: 
\begin{align}\label{eq:A_ansatz_1st_order}
    {\bf A}'({\bf r},t)^{(1)} = {\bf a}_0({\bf r}) + {\bf a}_0^*&({\bf r}) + {\bf a}_1({\bf r})e^{-i\omega t} + {\bf a}_1^*({\bf r})e^{i\omega t} \nonumber\\
    &+ {\bf a}_2({\bf r})e^{-i2\omega t} + {\bf a}_2^*({\bf r})e^{i2\omega t},
\end{align}
where $a_0 $ and $a_2$ refer to the  amplitudes for $0$ and $2\omega $ frequencies.  $|{\bf a}_1|\gg|{\bf a}_0|,|{\bf a}_2|$, and the superscript on LHS denotes  first-order nonlinear correction. We first consider the dynamics of $\delta\rho$. Since $\delta\rho$ appears only in the nonlinear term of Eq.\,(\ref{eq:Aprime_waveeq}), it is sufficient to use only the leading term in the ansatz of ${\bf A}'$ in the equation for $\rho$. By doing so we obtain
\begin{align}\label{eq:rho_1st_order}
\delta\rho({\bf r},t) =  \frac{1}{i\omega\mu q\lambda_L^2}&\big(\!-\!{\bf \nabla} \cdot {\bf a}_1e^{-i\omega t}  + {\bf \nabla} \cdot {\bf a}_1^*e^{i\omega t}\big).
\end{align}
Using Eq.\,(\ref{eq:rho_1st_order}) and the ansatz in Eq.\,(\ref{eq:A_ansatz_1st_order}), along with keeping track of up to only first order corrections, we can decompose Eq.\,(\ref{eq:Aprime_waveeq}) into separate equations for ${\bf a}_0, {\bf a}_1$ and ${\bf a}_2$ by grouping together all terms with the same harmonic order
\begin{equation}\label{eq:homogenNorm_a0}
     {\bf \nabla}\!\times\!{\bf \nabla}\!\times\!{\bf a}_0 -\frac{1}{\lambda_L^2}{\bf a}_0 -\frac{iq}{m\lambda_L^2\omega}({\bf \nabla} \cdot {\bf a}_1){\bf a}_1^* = 0,
\end{equation}
\begin{equation}\label{eq:homogenNorm_a1}
    {\bf \nabla}\!\times\!{\bf \nabla}\!\times\!{\bf a}_1 +  \bigg(\mu\epsilon \omega^2 -\frac{1}{\lambda_L^2}\bigg){\bf a}_1 = 0,
\end{equation}
and
\begin{align}\label{eq:homogenNorm_a2}
   {\bf \nabla}\!\times\!{\bf \nabla}\!\times\!{\bf a}_2 + \bigg(4\mu\epsilon\omega^2 - &\frac{1}{\lambda_L^2} \bigg){\bf a}_2 - \frac{iq}{m\lambda_L^2\omega}({\bf \nabla} \cdot {\bf a}_1){\bf a}_1 \nonumber\\
   &- i\mu\epsilon\frac{q}{m}\omega\nabla({\bf a}_1  \cdot {\bf a}_1) = 0.
\end{align}
In Eqs.\,(\ref{eq:homogenNorm_a0})-(\ref{eq:homogenNorm_a2}) above, we have considered the case where the density of normal electrons is negligible in the system, so that $\sigma=0$. Taking the divergence on both sides of Eq.\,(\ref{eq:homogenNorm_a1}), we get ${\bf \nabla} \cdot {\bf a}_1=0$. This is consistent with the fact that ${\bf a}_1$ is the solution of the equation for ${\bf A}'$ when only linear terms are considered. In such case we have $\delta\rho=0$, which is consistent with Eq.\,(\ref{eq:chargeconserve2}) for when the divergence of ${\bf A}'$ vanishes. Eq.\,(\ref{eq:homogenNorm_a1}) can then be simplified to
\begin{equation}\label{eq:homogenNorm_a1_1st_order}
    \nabla^2{\bf a}_1 +  \bigg(\mu\epsilon \omega^2 -\frac{1}{\lambda_L^2}\bigg){\bf a}_1 = 0,
\end{equation}
while the divergence term in Eq.\,(\ref{eq:homogenNorm_a0}) and Eq.\,(\ref{eq:homogenNorm_a2}) vanishes. 

To have the first correction to Eq.\,(\ref{eq:homogenNorm_a1}), the laplacian term in Eq.\,(\ref{eq:theta_EOM2}), which was neglected in the derivation of Eq.\,(\ref{eq:Aprime_waveeq}) needs to be considered. Taking this term into account, we obtain
\begin{equation}\label{eq:homogenNorm_a1_1st_order_full}
    {\bf \nabla}\!\times\!{\bf \nabla}\!\times{\bf a}_1 +  \bigg(\!\mu\epsilon \omega^2 -\frac{1}{\lambda_L^2}\!\bigg){\bf a}_1  + \frac{\mu\epsilon\hbar^2}{4m^2}\nabla\big(\nabla^2({\bf \nabla} \cdot {\bf a}_1)\big)= 0.
\end{equation}
Unlike Eq.\,(\ref{eq:homogenNorm_a1_1st_order}), Eq.\,(\ref{eq:homogenNorm_a1_1st_order_full}) does not lead to a divergence-free ${\bf a}_1$, and the charge fluctuation $\delta\rho$ is nonzero in this case. Taking the divergence on both side of Eq.\,(\ref{eq:homogenNorm_a1_1st_order_full}) we obtain an equation for $g({\bf r})={\bf \nabla} \cdot {\bf a}_1$ 
\begin{equation}\label{eq:div_a1_equation}
    \bigg(\mu\epsilon \omega^2 -\frac{1}{\lambda_L^2}\bigg)g + \frac{\mu\epsilon\hbar^2}{4m^2}{\bf \nabla} \cdot \big(\nabla(\nabla^2g)\big)= 0.
\end{equation}
In the presence of a harmonic source $\mu{\bf J}_{\text{src}}e^{-i\omega t}$, one can solve this equation for $g$. Let $G({\bf k})$ be the Fourier transform of $g({\bf r})$, then a Fourier transform of Eq.\,(\ref{eq:div_a1_equation}), with source included, yields 
\begin{equation}
    \bigg(\mu\epsilon \omega^2 -\frac{1}{\lambda_L^2}\bigg)G({\bf k}) + \frac{\mu\epsilon\hbar^2}{4m^2}k^4 G({\bf k}) = \mu\mathcal{F}\{{\bf \nabla} \cdot {\bf J}_{\text{src}} \},
\end{equation}
which leads to
\begin{equation}
    g({\bf r}) = \int_{-\infty}^{\infty}\frac{\mu\mathcal{F}\{{\bf \nabla}.{\bf J} \}}{\mu\epsilon \omega^2 -\frac{1}{\lambda_L^2} + \frac{\mu\epsilon\hbar^2}{4m^2}k^4}e^{i{\bf k}.{\bf r}} d{\bf k}.
\end{equation}

To treat Eqs.\,(\ref{eq:Aprime_waveeq}) and (\ref{eq:chargeconserve2}) to second order, we first consider the dynamical equation for $\delta\rho$, with first-order correction included
\begin{align}
    \frac{\partial\delta\rho}{\partial t} &= \frac{1}{q}{\bf \nabla} \cdot \Bigg[\bigg(\frac{1}{\mu\lambda_L^2} + \frac{q^2}{m}\delta\rho \bigg){\bf A}'\Bigg] \\
    &=\frac{1}{q}{\bf \nabla} \cdot \Bigg\{\!\bigg[\!\frac{1}{\mu\lambda_L^2} \!+\! \frac{iq}{\omega\mu m \lambda_L^2}\big({\bf \nabla} \cdot {\bf a}_1e^{-i\omega t} \!-\!  {\bf \nabla} \cdot {\bf a}^*_1e^{i\omega t}\big)\! \bigg]\!{\bf A}'\!\Bigg\}, \nonumber
\end{align}
which leads to
\begin{widetext}
\begin{align}
    \delta\rho({\bf r},t) = &\Bigg\{ \frac{1}{q\mu\lambda_L^2}{\bf \nabla}\cdot\Big({\bf a} + {\bf a}^*_0\Big) - \frac{i}{\omega\mu m \lambda_L^2}\Big[ {\bf \nabla}.\Big(({\bf \nabla}.{\bf a}_1){\bf a}_1^*\Big) - {\bf \nabla}\cdot\Big(({\bf \nabla}\cdot{\bf a}^*_1){\bf a}_1\Big) \Big] \Bigg\}t \nonumber\\
    & +\frac{i}{2\omega q \mu\lambda_L^2}\big(2{\bf \nabla} \cdot {\bf a}_1e^{-i\omega t} -  2{\bf \nabla}\cdot{\bf a}^*_1e^{i\omega t} + {\bf \nabla} \cdot {\bf a}_2e^{-i2\omega t} -  {\bf \nabla} \cdot {\bf a}^*_2e^{i2\omega t}  \big).
\end{align}
With the form of $\delta\rho$ shown above, the term $\delta\rho{\bf A}'$ in Eq.\,(\ref{eq:Aprime_waveeq}), up to two lowest orders of correction, generates the extra terms $te^{\pm i\omega t}$ and $e^{\pm i3\omega t}$. This suggest the following form for ${\bf A}'$
\begin{align}\label{eq:2nd_order_ansatz}
{\bf A}'({\bf r},t) = {\bf a}_0({\bf r}) &+ {\bf a}_0^*({\bf r}) + {\bf a}_1({\bf r})e^{-i\omega t} + {\bf a}_1^*({\bf r})e^{i\omega t} + {\bf a}_2({\bf r})e^{-i2\omega t} + {\bf a}_2^*({\bf r})e^{i2\omega t} \nonumber\\
& + {\bf a}_3({\bf r})e^{-i3\omega t} + {\bf a}_3^*({\bf r})e^{i3\omega t} + {\bf b}_1({\bf r})te^{-i\omega t} + {\bf b}^*_1({\bf r})te^{i\omega t},
\end{align}
where $b_1, a_3\ll a_0, a_2 \ll a_1$. With the emergence of the $te^{\pm i\omega t}$ terms, the system is prevented from going into a steady state. Using the form in Eq.\,(\ref{eq:2nd_order_ansatz}) to rewrite Eqs.\,(\ref{eq:Aprime_waveeq}) and (\ref{eq:chargeconserve2}) we obtain the following equations
\begin{equation}\label{eq:homogenNorm_a0_2ndorder}
     {\bf \nabla}\!\times\!{\bf \nabla}\!\times\!{\bf a}_0 -\frac{1}{\lambda_L^2}{\bf a}_0 -\frac{iq}{m\lambda_L^2\omega}({\bf \nabla}\cdot{\bf a}_1){\bf a}_1^* = 0,
\end{equation}

\begin{align}\label{eq:homogenNorm_a1_2ndorder}
    {\bf \nabla}\!\times\!{\bf \nabla}\!\times\!{\bf a}_1 &+ \bigg(\mu\epsilon\omega^2 - \frac{1}{\lambda_L^2}\bigg){\bf a}_1 + i2\omega\mu\epsilon{\bf b}_1 - \frac{iq}{2m\lambda_L^2\omega}\bigg\{\big({\bf \nabla}\cdot{\bf a}_2\big){\bf a}_1  + \frac{q}{m\omega}\Big[{\bf \nabla}\cdot\big(({\bf \nabla}\cdot{\bf a}_1){\bf a}_1 \big)\Big]{\bf a}_1^* \nonumber\\
    &+ 2({\bf \nabla}\cdot{\bf a}_1)({\bf a}_0 + {\bf a}_0^*) - 2({\bf \nabla}\cdot{\bf a}_1^*){\bf a}_2 \bigg\} - i\omega\mu\epsilon\frac{q}{m}\big[({\bf a}_0 + {\bf a}_0^*){\bf a}_1 + {\bf a}_2\cdot{\bf a}_1^* \big] = 0,
\end{align}

\begin{equation}\label{eq:homogenNorm_a2_2ndorder}
    {\bf \nabla}\!\times\!{\bf \nabla}\!\times\!{\bf a}_2 + \bigg(4\mu\epsilon\omega^2 - \frac{1}{\lambda_L^2} \bigg){\bf a}_2 - \frac{iq}{m\lambda_L^2\omega}({\bf \nabla}\cdot{\bf a}_1){\bf a}_1 - i\omega\mu\epsilon\frac{q}{m}\nabla({\bf a}_1\cdot{\bf a}_1) = 0,
\end{equation}

\begin{align}\label{eq:homogenNorm_a3_2ndorder}
    {\bf \nabla}\!\times\!{\bf \nabla}\!\times\!{\bf a}_3 + \bigg(9\mu\epsilon\omega^2 - \frac{1}{\lambda_L^2} \bigg){\bf a}_3 - \frac{iq}{2m\lambda_L^2\omega}\bigg\{\big({\bf \nabla}\cdot{\bf a}_2\big){\bf a}_1  + \frac{q}{m\omega}\Big[{\bf \nabla}\cdot\big(({\bf \nabla}\cdot{\bf a}_1)&{\bf a}_1 \big)\Big]{\bf a}_1 + 2({\bf \nabla}\cdot{\bf a}_1){\bf a}_2\bigg\} \nonumber\\
    & - i3\omega\mu\epsilon\frac{q}{m}({\bf a}_2\cdot{\bf a}_1) = 0,
\end{align}

\begin{equation}\label{eq:homogenNorm_b1_2ndorder}
    {\bf \nabla}\!\times\!{\bf \nabla}\!\times\!{\bf b}_1 + \bigg(\mu\epsilon\omega^2 - \frac{1}{\lambda_L^2} \bigg){\bf b}_1 - \frac{q}{m\lambda_L^2}\big[{\bf \nabla}\cdot({\bf a}_0 + {\bf a}_0^*)\big]{\bf a}_1 - \frac{iq^2}{m^2\lambda_L^2\omega}\Big[{\bf \nabla}\big(({\bf \nabla}\cdot{\bf a}_1){\bf a}_1^* \big) - {\bf \nabla}\big(({\bf \nabla}\cdot{\bf a}_1^*){\bf a}_1 \big) \Big]{\bf a}_1 = 0.
\end{equation}
\end{widetext}

\bibliography{refs}
\end{document}